\documentclass[preprint,12pt,compress,numafflabel]{elsarticle}

\usepackage{amssymb}
\usepackage{amsmath}
\usepackage{bm}
\usepackage{float}
\usepackage{graphicx}
\usepackage{booktabs}
\usepackage{longtable}
\usepackage{array}
\usepackage{hyperref}
\usepackage{xcolor}
\usepackage{listings}

\journal{Computer Physics Communications}

\newcommand{\Tr}{\mathrm{Tr}}

\newcommand{\Loss}{\mathcal{L}}

\begin{document}

\begin{frontmatter}

\title{MANDALA: An E(3)-Equivariant Graph Neural Network Framework for Learning Electronic-Structure Operators with Observable Guidance}

\author[casus,hzdr,uwr]{Bartosz Brzoza}
\author[casus,hzdr,wut]{Wiktoria Szopa}
\author[casus,hzdr]{Zakaria Elabid}
\author[casus,hzdr]{Vincent Martinetto}
\author[casus,hzdr]{Varadarajan Rengaraj}
\author[hzdr]{Mani Lokamani}
\author[casus,hzdr]{Thomas D. Kühne}
\author[casus,hzdr]{Attila Cangi\corref{cor1}}
\ead{a.cangi@hzdr.de}
\cortext[cor1]{Corresponding author}

\address[casus]{Center for Advanced Systems Understanding, 02826 Görlitz, Germany}
\address[hzdr]{Helmholtz-Zentrum Dresden-Rossendorf, 01328 Dresden, Germany}
\address[uwr]{Institute of Computer Science, University of Wroc\l aw, 50-300 Wroc\l aw, Poland}
\address[wut]{Faculty of Physics, Warsaw University of Technology, 00-662 Warsaw, Poland}

\begin{abstract}
Electronic-structure calculations based on Kohn-Sham density functional theory remain indispensable in computational materials science and chemistry. Their computational cost, however, limits accessible system sizes and simulation times.
At the same time, conventional machine-learning interatomic potentials (MLIPs), which are becoming the workhorse of large-scale materials modeling, usually target only energies and forces. They therefore leave out the quantum-operator-level information required to reconstruct band structures, densities of states, spatial charge distributions, and other electronic observables.
\texttt{Mandala} fills this methodological gap.
It is a modular software framework for learning block-sparse electronic-structure matrices with E(3)-equivariant graph neural networks. The framework is built around a unified representation of atom-resolved Hamiltonian, overlap, and density matrices, together with reusable abstractions for basis conversion, sparse block handling, irreducible representation mapping, graph construction, model definition, and training. This design allows \texttt{Mandala} to support heterogeneous chemical compositions, a wide range of neural architecture variants within one workflow, and multiple electronic-structure backends.
\texttt{Mandala} evaluates selected observables directly from the predicted operators, including band energy, electron count, density of states, and band structure. This connects electronic-structure learning and observable-guided modeling while retaining a representation tied to quantum-mechanical operators rather than only scalar or vector targets as in MLIPs.
In this form, \texttt{Mandala} is intended to complement atomistic interatomic potential workflows by resolving electronic structure and operator-derived observables within one scalable implementation.

\end{abstract}

\begin{keyword}
Electronic structure \sep Density functional theory \sep Equivariant graph neural networks \sep Sparse matrix learning \sep Operator-derived observables \sep Scientific machine learning
\end{keyword}

\end{frontmatter}

{\bf PROGRAM SUMMARY}

\begin{small}
\noindent
{\em Program Title:} \texttt{Mandala} \\
{\em CPC Library link to program files:} [TODO: to be added by the Technical Editor] \\
{\em Developer's repository link:} \url{https://github.com/Mandala-org/Mandala} \\
{\em Licensing provisions:} Apache-2.0 \\
{\em Programming language:} Python \\
{\em Nature of problem:} Electronic-structure calculations based on Kohn-Sham density functional theory are the central computational method in computational materials science and chemistry, but their computational cost limits accessible system sizes, trajectory lengths, and accessed physical properties. Atomistic simulations based on machine-learned interatomic potentials alleviate this cost for energies and forces, yet energy-and-force MLIPs do not, by themselves, provide the operator-level information needed to analyze charge redistribution or electronic spectra. A complementary route is to learn sparse representations of electronic-structure matrices themselves. Such a computational method must support multiple atom types, changing orbital bases, periodic boundary conditions, equivariant geometric processing, and operator-derived observables. \\
{\em Solution method:} \texttt{Mandala} represents Hamiltonian, overlap, and density matrices as atom-pair-resolved sparse block tensors and maps these blocks to irreducible E(3) representations compatible with equivariant neural networks. The package combines parser modules for electronic-structure data, basis-conversion utilities, a block-irrep mapper, graph construction routines, and configurable E(3)-equivariant message-passing models. Multi-head prediction allows simultaneous learning of several matrices, while band-energy and electron-count observables are evaluated from the predicted operators using differentiable sparse traces and may be included as training objectives. For periodic systems, optional spectral guidance compares generalized-eigenvalue spectra derived from the predicted Hamiltonian and overlap operators with reference spectra on a selected k-point mesh.
The framework is organized to support extensibility, architecture search, and reproducible training workflows. \\
{\em Additional comments including restrictions and unusual features:} The framework is optimized for sparse matrix learning of electronic structure in atomistic systems with localized orbital bases and periodic or nonperiodic geometries. A distinctive feature of \texttt{Mandala} is the combination of operator-level learning with optional observable guidance derived from those operators, which allows the same software stack to serve both electronic-structure emulation and observable-guided atomistic applications. The package is designed to accommodate multiple data backends together with a broad search space of equivariant architectures and training objectives.
\end{small}

\section{Introduction}
\label{sec:introduction}

Kohn--Sham density functional theory (DFT) is the standard method of electronic-structure simulation because it offers a practical compromise between physical fidelity and computational cost across chemistry, condensed-matter physics, and materials science \cite{HoKo1964,KoSh1965}. Even so, both the unfavorable system-size scaling and repeated self-consistent solution of the Kohn--Sham equations remain major bottlenecks in workflows that require large supercells, long molecular-dynamics trajectories, broad compositional screening, or repeated re-evaluation under changing thermodynamic conditions. This cost has motivated sustained interest in machine-learned surrogates for atomistic and quantum-mechanical simulation \cite{deepdive}. Relevant directions include machine-learned interatomic potentials (MLIPs) for potential-energy surfaces and forces \cite{BehlerParrinello2007,BartokGAP2010,ZhangDeepMD2018,BatznerNequIP2022,BatatiaMACE2022}, wave-function-related quantities \cite{PfauFermiNet2020,HermannPauliNet2020}, learned electron densities \cite{BrVo2017,GrisafiElectronDensity2019,malapaper,Rackers_2023}, kinetic-energy functional learning for orbital-free DFT \cite{ALGHADEER2021127621,ghasemi2021}, and other reduced electronic representations such as the local density of states and reduced density matrices \cite{malapaper,Shao_2023,chandrasekaran_solving_2019,temperaturepaper,sizetransferpaper}.

The most widely used machine-learning approach in chemistry, condensed-matter physics, and materials science is to train MLIPs, which bypass explicit electronic-structure calculations and instead learn observables such as energies and forces. The strategy is highly effective, from early neural-network potentials to modern equivariant graph models \cite{BehlerParrinello2007,BartokGAP2010,ZhangDeepMD2018,BatznerNequIP2022,BatatiaMACE2022}. Recent work has shown that equivariant neural architectures are especially well suited for energy and force prediction because they can encode rotational structure directly in the model rather than relying entirely on handcrafted invariants or data augmentation \cite{KoKoBe22,BatznerMusaelianKozinskyNRP2023,KoOngNatCompSci2023}. However, learning only energies and forces limits direct access to the electronic structure itself, including operator-valued quantities and Hamiltonian-derived analyses such as band structure. For applications that remain fundamentally electronic-structure-centered, it is therefore attractive to learn the operators themselves.
Examples include systems in which charge transfer, orbital hybridization, defect-localized states, response to external fields, or band-structure-level analysis are central to the scientific question. In such settings, access to the underlying electronic operators is not merely auxiliary metadata but part of the modeling target itself.

\texttt{Mandala} is designed around this operator-learning viewpoint. The framework targets block-sparse matrix quantities arising in localized-orbital electronic-structure calculations, in particular the Hamiltonian, overlap, and density matrices. Instead of treating these objects as dense global arrays, \texttt{Mandala} represents them as atom-pair-resolved sparse blocks and combines this representation with E(3)-equivariant graph neural networks. The resulting learning problem respects both the locality of the underlying electronic structure and the geometric transformation laws imposed by three-dimensional space.

The development of \texttt{Mandala} is motivated by three specific design goals. First, it should be modular at the level of both scientific abstractions and code organization. Data ingestion, basis conversion, graph construction, irrep mapping, model definition, and training logic should be independently replaceable while still composing into a single workflow. Second, the framework should support supervision beyond raw matrix reconstruction. In addition to differentiable sparse-trace observables, periodic Hamiltonian and overlap predictions can optionally be guided by a spectral loss that compares generalized-eigenvalue information on a selected k-point mesh. When Hamiltonian, overlap, and density matrices are predicted jointly, the band-energy observable and electron count can be computed directly from the predicted operators and included in the training objective. Third, the framework should make architecture exploration an inherent capability rather than an afterthought. Choices such as irreducible feature content, message-passing depth, tensor-product design, head structure, residual connections, normalization, and observable losses must all be accessible through a coherent configuration and sweep interface.

Related operator-learning efforts include the DeepH family of Hamiltonian-learning models for localized-basis DFT, including their original message-passing formulation, extensions to magnetic superstructures, an explicitly E(3)-equivariant formulation, hybrid-functional Hamiltonians, and conversion pathways from plane-wave data to localized-orbital Hamiltonians \cite{deeph2022,deeph_magnetic_2023,deephe3_2023,deeph_hybrid_2024,deeph_pw_2024,deeph_review_2025}. Closely related is MACE-H, which combines many-body message passing with equivariant processing for Hamiltonian prediction \cite{maceh_2026}. These works demonstrate the promise of Hamiltonian learning at scale, while our emphasis here is on a reusable software framework that unifies sparse operator prediction, observable guidance, multi-backend data handling, and broad architecture exploration.

These goals distinguish \texttt{Mandala} from earlier machine-learning workflows in two ways. Relative to purely observable-driven MLIPs, \texttt{Mandala} preserves an operator-level representation that remains useful for electronic-structure analysis. Relative to monolithic research scripts built around one specific dataset or one fixed neural architecture, it provides a reusable framework in which data formats, physical targets, and model variants can be exchanged without rewriting the entire code path.

\begin{figure}[H]
\centering
\includegraphics[width=\textwidth]{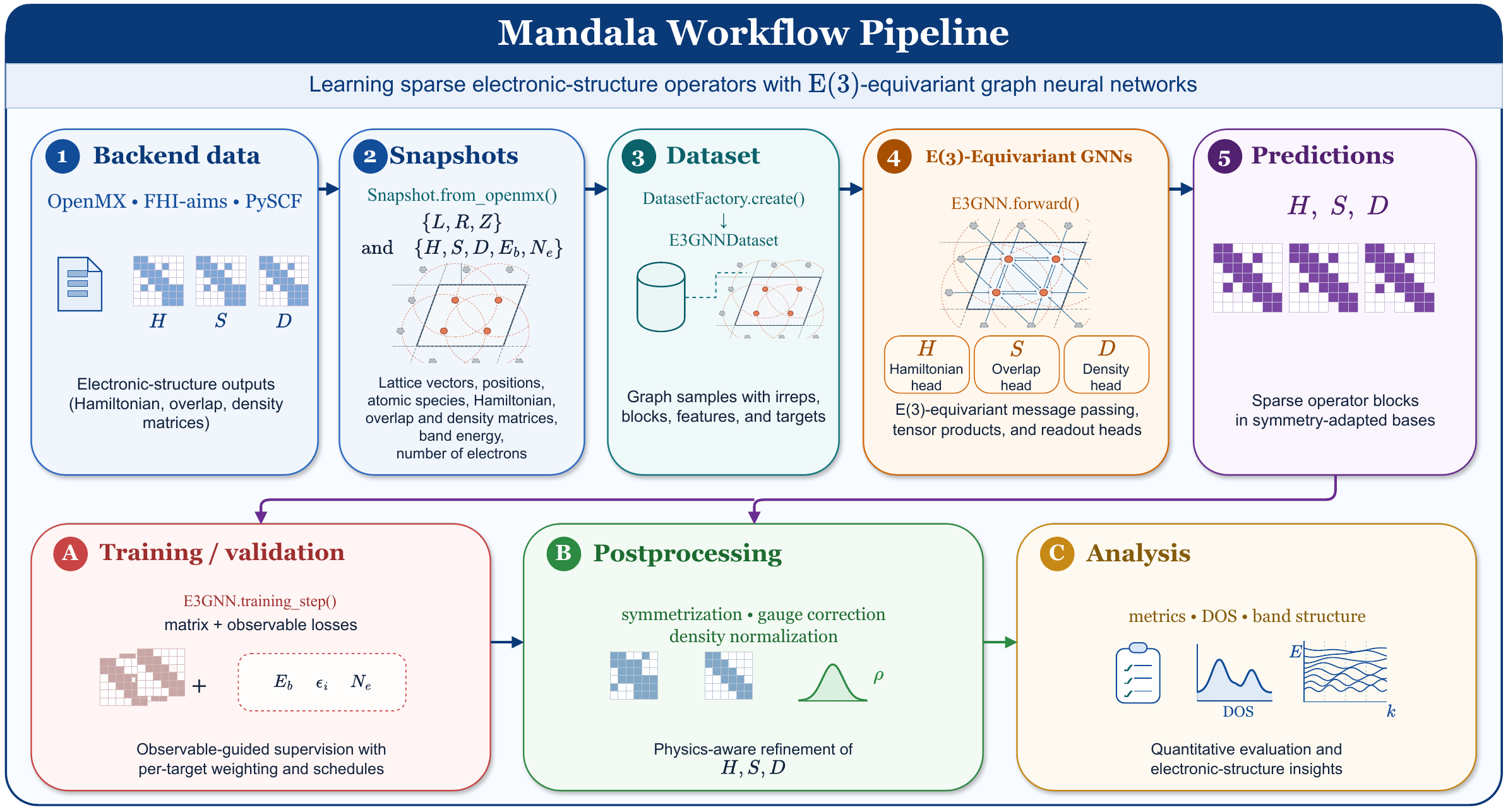}
\caption{End-to-end \texttt{Mandala} workflow. Outputs from localized-orbital electronic-structure backends are standardized as \texttt{Snapshot} objects and assembled into graph samples with sparse operator targets. An E(3)-equivariant graph neural network maps these samples to Hamiltonian, overlap, and density blocks. The predicted operators support direct matrix and observable supervision, numerical safeguards and normalization, and subsequent evaluation through matrix diagnostics, densities of states, and band structures.}
\label{fig:workflow_overview}
\end{figure}

Figure~\ref{fig:workflow_overview} provides an overview of this operator-centered workflow:
electronic-structure outputs are first converted into backend-independent snapshots, from which \texttt{Mandala} constructs graph samples with a common sparse-block and symmetry convention.
An E(3)-equivariant model then predicts Hamiltonian, overlap, and density blocks in symmetry-adapted bases.
The same predicted operators connect model fitting to physics-aware postprocessing and to electronic-structure analyses, so that matrix supervision, observable guidance, and quantities such as densities of states and band structures remain part of one consistent pipeline.

The remainder of this paper is organized as follows: section~\ref{sec:background} provides the necessary theoretical background. Section~\ref{sec:implementation} describes the sparse-matrix data model, the E(3)-equivariant learning formulation, the observable-aware loss construction, and the software abstractions. Section~\ref{sec:results} showcases \texttt{Mandala}'s capabilities by predicting the Hamiltonian matrix, density matrix, density of states, and band structure for selected systems, including the chalcogenide ZnCu$_2$Sn(SeS)$_2$, amorphous silicon oxide, and crystalline silicon. Section~\ref{sec:ablations} presents an ablation study, and the Appendix provides additional model details.

\section{Theoretical Background}
\label{sec:background}

\subsection{From the full electron-ion Schr\"odinger equation to Kohn-Sham DFT}
\label{subsec:dft_background}

At the most fundamental nonrelativistic level, an atomistic system is described by the stationary many-body electronic Schr\"odinger equation for all electrons and all nuclei within the Born-Oppenheimer approximation~\cite{born_zur_1927},
\begin{equation}
\left[ T_e(\underline{\bm{r}}) + V_{ee}(\underline{\bm{r}}) + V_{ei}(\underline{\bm{r}};\underline{\bm{R}}) + E_{ii}(\underline{\bm{R}}) \right] \Psi(\underline{\bm{r}};\underline{\bm{R}})
=
E \Psi(\underline{\bm{r}};\underline{\bm{R}}),
\end{equation}
where $\Psi(\underline{\bm{r}};\underline{\bm{R}})$ denotes the many-body wavefunction, $\underline{\bm{r}}$ the collection of electronic coordinates, and $\underline{\bm{R}}$ the collection of ionic coordinates. The semicolon indicates that the dependence on ionic coordinates is parametric.
The electronic Schrödinger equation is composed of the kinetic energy of the electrons
\begin{equation}
T_e(\underline{\bm{r}}) = -\frac{1}{2}\sum_{j=1}^{N_e} \nabla_j^2,
\end{equation}
the electron-electron interaction
\begin{equation}
V_{ee}(\underline{\bm{r}}) =
\frac{1}{2}\sum_{j=1}^{N_e}\sum_{k \ne j}^{N_e}\frac{1}{|\bm{r}_j-\bm{r}_k|},
\end{equation}
the electron-ion interaction
\begin{equation}
V_{ei}(\underline{\bm{r}};\underline{\bm{R}}) =
-\sum_{j=1}^{N_e}\sum_{\alpha=1}^{N_i}\frac{Z_{\alpha}}{|\bm{r}_j-\bm{R}_{\alpha}|},
\end{equation}
and the ion-ion interaction
\begin{equation}
E_{ii}(\underline{\bm{R}}) =
\frac{1}{2}\sum_{\alpha=1}^{N_i}\sum_{\beta \ne \alpha}^{N_i}
\frac{Z_{\alpha} Z_{\beta}}{|\bm{R}_{\alpha}-\bm{R}_{\beta}|}.
\end{equation}
Within the Born-Oppenheimer approximation, the ion-ion interaction amounts to a constant energy shift for fixed ionic configuration, while the other terms depend only parametrically on $\underline{\bm{R}}$. In the following, this parametric dependence is not always written explicitly. Also note that we adopt atomic units throughout, with $\hbar = m_e = e^2 = 1$, so that energies are measured in Hartree and lengths in Bohr radii.

The central task in electronic structure theory is therefore solving the interacting many-electron problem defined by $T_e + V_{ee} + V_{ei}$ for a fixed ionic geometry.

DFT is the most popular method for solving this problem. It replaces the many-electron wavefunction by the electron density $n(\bm{r})$ as the central variable \cite{HoKo1964}. In the Kohn-Sham construction \cite{KoSh1965}, the interacting electron problem is then mapped onto a fictitious system of non-interacting electrons governed by the Kohn-Sham equations
\begin{equation}
H \psi_j(\bm{r}) = \epsilon_j \psi_j(\bm{r}),
\end{equation}
where $H = -\frac{1}{2}\nabla^2 + v_S(\bm{r})$ denotes the Kohn-Sham Hamiltonian and $j=1,\dots,N_e$ labels the Kohn-Sham orbitals. Solving the Kohn-Sham equations yields the corresponding electronic density
\begin{equation}
n(\bm{r}) = \sum_{j=1}^{N_e} |\psi_j(\bm{r})|^2
\end{equation}
which equals the electronic density of the interacting many-body system. The equality of both densities is achieved by the Kohn-Sham potential
\begin{equation}
v_S(\bm{r}) =
\frac{\delta E_H[n]}{\delta n(\bm{r})}
+ \frac{\delta E_{XC}[n]}{\delta n(\bm{r})}
+ v_{ei}(\bm{r}),
\end{equation}
where $E_H[n]$ is the Hartree energy, $E_{XC}[n]$ the exchange-correlation energy, and
\begin{equation}
v_{ei}(\bm{r}) = -\sum_{\alpha} \frac{Z_{\alpha}}{|\bm{r}-\bm{R}_{\alpha}|}
\end{equation}
the external potential. Although the Kohn--Sham construction is formally exact, practical calculations rely on approximations to the exchange-correlation functional.
In this notation, the ground-state energy is written as
\begin{equation}
E[n] =
T_S[n] + E_H[n] + E_{XC}[n]
+ \int d\bm{r}\, n(\bm{r}) v_{ei}(\bm{r})
+ E_{ii},
\end{equation}
where $T_S[n]$ is the Kohn-Sham kinetic energy.

\texttt{Mandala} connects to the electronic structure problem at the level of Kohn-Sham DFT and operates on localized-orbital matrix representations of this problem.
Expanding the Kohn-Sham orbitals in a finite basis $\{\chi_{\mu}\}$,
\begin{equation}
\psi_j(\bm{r}) = \sum_{\mu} C_{\mu j}\chi_{\mu}(\bm{r}),
\end{equation}
leads to the generalized eigenvalue problem
\begin{equation}
\sum_{\nu} H_{\mu \nu} C_{\nu j}
=
\epsilon_j \sum_{\nu} S_{\mu \nu} C_{\nu j},
\end{equation}
with
\begin{equation}
H_{\mu \nu} = \langle \chi_{\mu} | H | \chi_{\nu} \rangle,
\qquad
S_{\mu \nu} = \langle \chi_{\mu} | \chi_{\nu} \rangle.
\end{equation}
The one-particle density matrix is
\begin{equation}
D_{\mu \nu} = \sum_j f_j C_{\mu j} C_{\nu j}^{*},
\end{equation}
where $f_j$ denotes the orbital occupation. In a nonorthogonal basis, the electron count is
\begin{equation}
N_e = \Tr(DS),
\end{equation}
while the band energy is obtained from
\begin{equation}
E_{\mathrm{b}} = \Tr(DH).
\end{equation}
In this work, \texttt{Mandala} uses the trace expressions above as operator-derived observables associated with the predicted matrices.

Localized basis sets also induce block sparsity. If orbital $\mu$ belongs to atom $i$ and orbital $\nu$ belongs to atom $j$, then matrix entries can be organized into atom-pair blocks $X_{ij}^{\bm{L}}$, where $X \in \{H,S,D\}$ and $\bm{L} \in \mathbb{Z}^3$ indexes periodic lattice images. This sparse block structure is the operator domain learned by \texttt{Mandala}.

\subsubsection{Hamiltonian gauge invariance and the overlap-shift correction}
\label{subsubsec:gauge_invariance}

In a nonorthogonal basis, the generalized eigenvalue problem is invariant under a rigid shift of the Hamiltonian by a multiple of the overlap matrix. Given the generalized eigenvalue equation,
\begin{equation}
H C = S C \varepsilon,
\end{equation}
we can define a shifted Hamiltonian:
\begin{equation}
\widetilde{H} = H - \alpha S.
\end{equation}
Applying this to the eigenvectors yields:
\begin{equation}
\widetilde{H} C = (H-\alpha S) C = S C \varepsilon - \alpha S C = S C (\varepsilon - \alpha I).
\end{equation}
Therefore, the generalized eigenvectors remain unchanged, while all eigenvalues are uniformly shifted by the constant $\alpha$. This demonstrates that setting the energy reference point can be represented in matrix form by the transformation $H \mapsto H - \alpha S$.

Furthermore, for a fixed number of electrons ($N_e$) and a geometry-independent shift $\alpha$, the expectation value of the energy evaluates to:
\begin{equation}
\operatorname{Tr}(D\widetilde{H}) = \operatorname{Tr}(DH) - \alpha \operatorname{Tr}(DS) = \operatorname{Tr}(DH) - \alpha N_e.
\end{equation}
For fixed electron number, this transformation preserves the generalized eigenvectors and shifts all one-electron eigenvalues and the band energy by a constant reference offset.

For error reporting, \texttt{Mandala} accounts for this gauge freedom through the overlap-shift alignment $\mu_H$, which is chosen to minimize the mean-squared discrepancy between a predicted Hamiltonian and a reference Hamiltonian after subtraction of a multiple of the overlap matrix:
\begin{equation}
\mu_H
=
\arg\min_{\mu}
\left\|
H^{\mathrm{pred}} - H^{\mathrm{ref}} - \mu S
\right\|_F^2.
\end{equation}
Taking the derivative with respect to $\mu$ and setting it to zero yields
\begin{equation}
\mu_H
=
\frac{
\left\langle
H^{\mathrm{pred}} - H^{\mathrm{ref}}, S
\right\rangle_F
}{
\left\langle S, S \right\rangle_F
},
\end{equation}
where $\langle A,B\rangle_F = \sum_{\mu \nu} A_{\mu \nu} B_{\mu \nu}$ is the Frobenius inner product, evaluated in \texttt{Mandala} blockwise over the aligned sparse representation. Because $\mu_H$ depends on $H^{\mathrm{ref}}$, this is a reference-dependent alignment used only for fair matrix-error reporting: two Hamiltonians that differ by $cS$ produce the same wavefunctions and spectra separated only by the arbitrary constant $c$. It is not an inference-time correction. In deployment, the learned Hamiltonian retains the energy reference of the training backend; spectral comparisons instead report energies relative to the reference Fermi level.

\subsection{Scope, conventions, and assumptions}
\label{subsec:scope-conventions}

The present implementation operates on localized atomic-orbital bases with explicit orbital and shell metadata for each species. The demonstrations use real, non-spin-polarized matrices without spin--orbit coupling. Periodic and nonperiodic structures are supported by the sparse data model; the demonstrations and spectral objective in this work are periodic. Geometry is represented in \r{A}, cutoff radii are specified in \r{A}, Hamiltonians are stored internally in Hartree, and reported Hamiltonian and spectral errors are converted to electronvolts. Overlap and density matrices are dimensionless in the adopted basis convention. The reference-dependent Hamiltonian gauge alignment is restricted to error reporting.

For a lattice translation $\bm{L}$ and fractional reciprocal coordinate $\bm{k}$, the periodic Fourier convention is
\begin{equation}
X(\bm{k})=\sum_{\bm{L}}\exp\!\left(i\bm{k}\cdot\bm{L}\right)X^{\bm{L}},
\qquad X\in\{H,S,D\}.
\end{equation}
Backend-specific real spherical-harmonic conventions and orbital orderings are converted to a common e3nn convention during snapshot preparation. The block-to-irrep transform therefore depends on the complete orbital and shell metadata of both atoms, rather than on chemical species alone. Matrix targets and graph edges are filtered to the configured real-space cutoff, including the required periodic images.

\subsection{E(3) symmetry, equivariance, and irreducible representations}
\label{subsec:egnn_background}

Atomistic systems in three-dimensional space are naturally acted on by the Euclidean group $\mathrm{E}(3)=\mathbb{R}^3 \rtimes \mathrm{O}(3)$, combining translations, rotations, and reflections.
For a group element $g=(Q,\bm{t})$, with $Q \in \mathrm{O}(3)$ and $\bm{t}\in\mathbb{R}^3$, atomic positions transform as
\begin{equation}
\bm{R}_i \mapsto g \cdot \bm{R}_i = Q \bm{R}_i + \bm{t}.
\end{equation}
A model output $f(X)$ is invariant if $f(g \cdot X)=f(X)$, and equivariant if
\begin{equation}
f(g \cdot X) = \rho_{\mathrm{out}}(g)\, f(X),
\end{equation}
for a representation $\rho_{\mathrm{out}}$ appropriate to the predicted object.
Scalar observables such as total energies are invariant. Vectors such as forces transform covariantly, and operator-valued quantities inherit structured transformation rules from the orbital basis in which they are expressed.

In equivariant neural networks, hidden features are not treated as generic channels. Instead, they are decomposed into irreducible representations (irreps) of $\mathrm{O}(3)$, labeled by angular momentum $\ell$ and parity $p \in \{+1,-1\}$. A feature space therefore takes the form
\begin{equation}
\mathcal{H}
=
\bigoplus_{\ell,p} m_{\ell,p} \, D^{(\ell,p)},
\end{equation}
where $m_{\ell,p}$ is the multiplicity of irrep $(\ell,p)$ and
$D^{(\ell,p)}$ is the corresponding representation matrix.
Scalars correspond to $(\ell=0,p=+1)$, ordinary vectors to $(\ell=1,p=-1)$, and higher-rank objects are built from larger $\ell$ channels.
This representation-theoretic bookkeeping is particularly natural for localized atomic orbitals, whose angular content is already organized by orbital angular momentum.

The key architectural consequence is that every learned operation must respect the decomposition into irreps.
Node and edge features are updated through maps that commute with the group action, rather than by unrestricted dense linear
layers.
Geometric information enters through relative displacement vectors and their spherical-harmonic expansions, while translational invariance is enforced by expressing all geometry in relative coordinates.
Reflections are tracked through the parity label $p$, while rotations act within each $\ell$ channel.
In this way, \texttt{Mandala} preserves directional information throughout the network without sacrificing the required covariance laws. The implementation uses the \texttt{e3nn} library for irreducible representations, spherical harmonics, and equivariant tensor-product operations~\cite{GeigerSmidt2022}.

\subsection{Tensor products, spherical harmonics, and the Wigner-Eckart viewpoint}
\label{subsec:tensor_products_background}

The basic algebraic operation of an equivariant graph neural network is the tensor product of irreps.
If two feature channels transform according to
$D^{(\ell_1,p_1)}$ and $D^{(\ell_2,p_2)}$, then their tensor product decomposes as
\begin{equation}
D^{(\ell_1,p_1)} \otimes D^{(\ell_2,p_2)}
=
\bigoplus_{L=|\ell_1-\ell_2|}^{\ell_1+\ell_2}
D^{(L,p_1 p_2)}.
\end{equation}

This Clebsch-Gordan structure determines which output irreps may be produced by combining two input channels.
In practical terms, it is the algebraic rule that governs message passing, nonlinear mixing of hidden features, and the final projection from latent features to matrix coefficients.

Edge geometry is injected through spherical harmonics $Y_{\ell m}(\hat{\bm{r}}_{ij})$, evaluated on normalized relative displacement
directions.
Because spherical harmonics transform irreducibly under rotations, they provide the canonical angular basis for equivariant edge messages.
A typical equivariant interaction therefore combines learned node or edge features with spherical-harmonic features through tensor products, while radial functions modulate the interaction strength as a function of distance.

The same representation-theoretic structure appears in matrix elements of operators between angular-momentum-carrying orbitals. For a spherical tensor operator $T^{(k)}_q$, the Wigner-Eckart theorem states that
\begin{equation}
\langle \alpha \ell m | T^{(k)}_q | \alpha' \ell' m' \rangle
=
\langle \ell' m'; k q | \ell m \rangle
\langle \alpha \ell \| T^{(k)} \| \alpha' \ell' \rangle,
\end{equation}
where $\ell$ and $\ell'$ denote orbital angular momenta, $m$ and $m'$ their magnetic quantum numbers, $T^{(k)}_q$ is the $q$th component of a spherical tensor operator of rank $k$, $\alpha$ and $\alpha'$ collect any additional quantum numbers needed to specify the states, and $\langle \ell' m'; k q | \ell m \rangle$ is a Clebsch-Gordan coefficient.
The double-bar matrix element $\langle \alpha \ell \| T^{(k)} \| \alpha' \ell' \rangle$ is the corresponding reduced matrix element.
In other words, the Clebsch--Gordan coefficient fixes the purely geometric way that angular components couple under rotations, while the reduced matrix element is the remaining rotationally invariant coefficient.
It may depend on atomic species, radial basis functions, interatomic distance, and the local chemical environment, but not on the choice of coordinate axes.

\texttt{Mandala} uses this viewpoint operationally: the angular coupling structure is fixed analytically by the irrep mapping, while the neural network learns the reduced coefficients associated with each orbital pair and chemical environment.
This is precisely why the matrix-block prediction problem can be posed naturally in an irrep basis rather than directly in Cartesian matrix entries.

\subsection{From equivariant local features to sparse operator prediction}
\label{subsec:matrix_observable_background}

\texttt{Mandala} represents each operator $X \in \{H,S,D\}$ as a sparse family of atom-pair blocks $X_{ij}^{\bm{L}}$, where $i$ and $j$ index atoms and $\bm{L}$ labels periodic images.
The block representation is converted into an irrep representation by a fixed change of basis,
\begin{equation}
\bm{x}_{ij}^{\bm{L}} = Q_{a_i a_j} \, \mathrm{vec}\!\left(X_{ij}^{\bm{L}}\right),
\end{equation}
where $a_i$ and $a_j$ denote the atomic species of atoms $i$ and $j$, and $Q_{a_i a_j}$ is the species-pair-specific block-to-irrep transform.
The inverse map reconstructs matrix blocks from predicted irrep vectors,
\begin{equation}
\mathrm{vec}\!\left(X_{ij}^{\bm{L}}\right)
=
Q_{a_i a_j}^{-1} \bm{x}_{ij}^{\bm{L}},
\end{equation}
which, in the orthonormal convention used internally, is a transpose.
In the code these maps are materialized by the \texttt{BlockIrrepMapper}. The same component is reused by the data pipeline, the model heads, and the evaluation routines.

Once the network predicts a consistent set of sparse operators, selected quantities are obtained directly from those operators rather than through a separate surrogate model.
For sparse block matrices, the trace of a product is evaluated as a sparse contraction
\begin{equation}
\Tr(AB)
=
\sum_{(\bm{L},i,j)}
\mathrm{tr}\!\left(
A_{ij}^{\bm{L}} B_{ji}^{-\bm{L}}
\right),
\end{equation}
implemented in \texttt{Mandala}. This gives direct access to electron counts and band-energy observables from the same predicted operator set.
In this way, internal consistency between matrices and observables is built into the model directly.

\section{Numerical Implementation}
\label{sec:implementation}

\subsection{Software design goals, sparse data model, and periodic graph construction}
\label{subsec:workflow}

\texttt{Mandala} is organized as a layered scientific software stack whose core workflow consists of parsing electronic-structure outputs, harmonizing basis conventions, building sparse operators and graphs, predicting irrep vectors, reconstructing operators, and evaluating observables.

\begin{figure}[H]
\centering
\includegraphics[width=\textwidth]{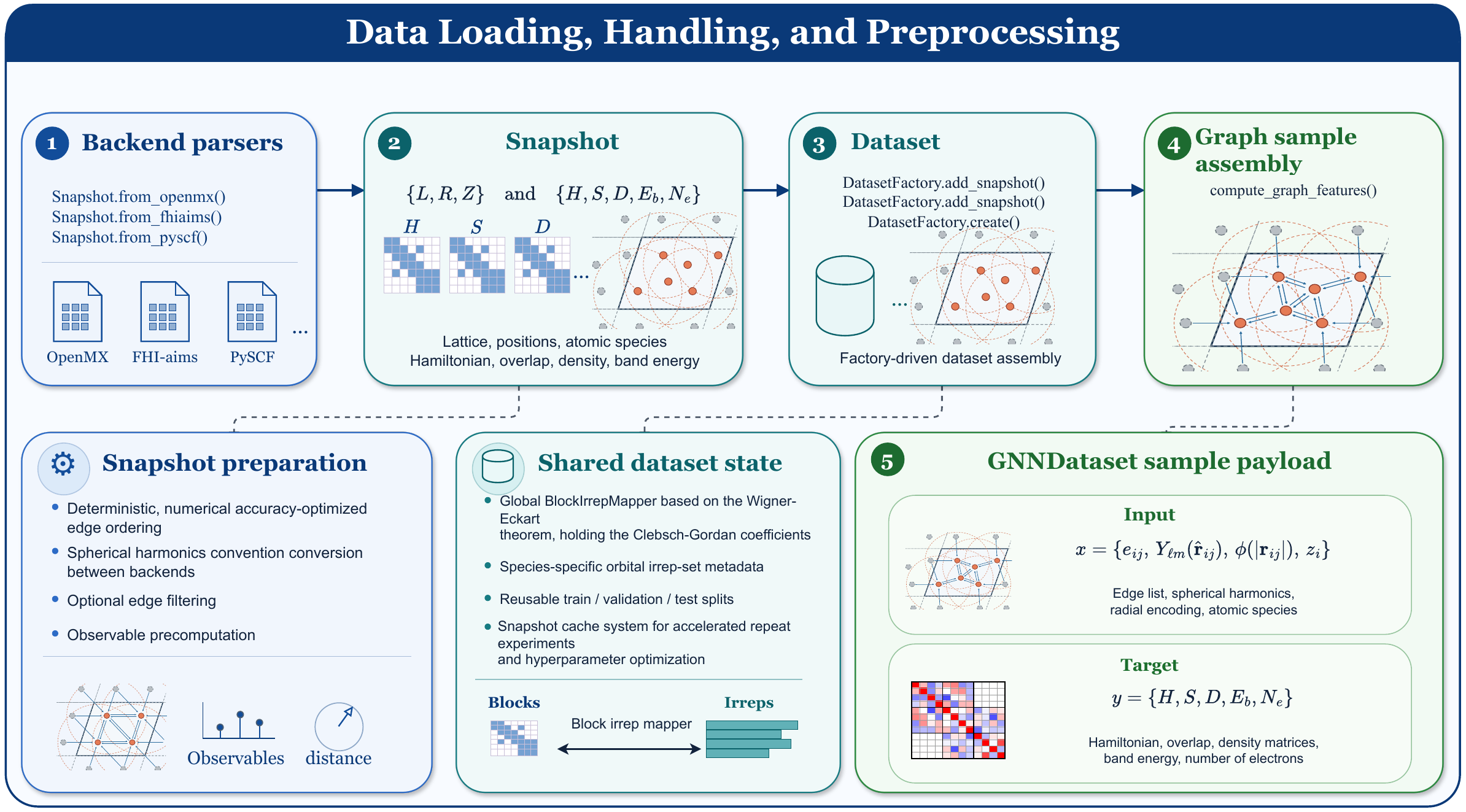}
\caption{Data loading and graph-sample construction. Backend-specific parsers convert OpenMX, FHI-aims, and PySCF electronic-structure outputs into a common \texttt{Snapshot} containing the atomic structure, sparse Hamiltonian, overlap, and density blocks, and available scalar observables. \texttt{DatasetFactory} combines snapshots while sharing orbital metadata, a global \texttt{BlockIrrepMapper}, reusable train/validation/test splits, and cached preprocessing state. Graph assembly establishes deterministic periodic edges, harmonizes backend conventions, and computes radial and spherical-harmonic features. The resulting \texttt{E3GNNDataset} sample pairs these graph inputs with matrix and observable targets in a consistent indexing and symmetry convention.}
\label{fig:data_preprocessing}
\end{figure}

Figure~\ref{fig:data_preprocessing} summarizes the data-loading and sample-construction path.
The central data abstractions are \texttt{BlockMatrix}, \texttt{IrrepsBlockData}, and \texttt{Snapshot}.
A \texttt{Snapshot} contains all matrices and metadata associated with one geometry, including positions and the simulation cell.
This keeps Hamiltonian, overlap, and density predictions tied to a single physically consistent basis and sparse indexing scheme.

The sparse operator representation stores blocks by ordered species pair and by explicit edge metadata $(\bm{L},i,j)$, where $\bm{L}\in\mathbb{Z}^3$ is the lattice-image shift.
Basis-conversion modules map backend-specific orbital orderings and sign conventions into the common internal representation expected by the equivariant stack.
In the current framework, OpenMX~\cite{ozaki2003,ozakikino2004}, FHI-aims~\cite{blum2009}, and PySCF~\cite{pyscf2020} parsers and convention mappers are available.

Graph construction mirrors the sparse operator indexing. Nodes correspond to atoms, while edges correspond to self-interactions and neighbor interactions inside a cutoff radius.
For periodic systems, each edge carries an integer shift vector $\bm{L}$ and a relative displacement
\begin{equation}
\bm{r}_{ij}^{\bm{L}}
=
\bm{R}_j - \bm{R}_i + \sum_{\alpha=1}^{3} L_{\alpha} \bm{a}_{\alpha},
\end{equation}
where $\{\bm{a}_{\alpha}\}$ are the cell vectors.
These displacements are used to construct radial embeddings and spherical-harmonic features.
The directed periodic edge set is therefore
\begin{equation}
\mathcal{E}
=
\left\{
(i,j,\bm{L})
\;\middle|\;
\left\|
\bm{r}_{ij}^{\bm{L}}
\right\|_2
\le r_{\mathrm{c}}
\right\},
\end{equation}
with a further partition into diagonal, shifted-self, and off-diagonal edges,
\begin{align}
\mathcal{E}_{\mathrm{diag}}
&=
\left\{
(i,j,\bm{L}) \in \mathcal{E}
\;\middle|\;
i=j,\;
\bm{L}=\bm{0}
\right\}, \\
\mathcal{E}_{\mathrm{shifted\_self}}
&=
\left\{
(i,j,\bm{L}) \in \mathcal{E}
\;\middle|\;
i=j,\;
\bm{L}\neq \bm{0}
\right\}, \\
\mathcal{E}_{\mathrm{offdiag}}
&=
\left\{
(i,j,\bm{L}) \in \mathcal{E}
\;\middle|\;
i\neq j
\right\}.
\end{align}
To preserve exact agreement between graph edges and sparse matrix blocks, \texttt{Mandala} imposes a canonical ordering of edges by distance and lattice shift, with reverse-edge lookup tables used for sparse traces and symmetrization.
This explicit reverse alignment is essential for both physical correctness and reproducibility under periodic boundary conditions.

The same snapshot-level representation preserves access to operator-derived postprocessing.
In particular, the predicted Hamiltonian may be used for band-structure analysis within the same sparse-operator workflow through \texttt{Snapshot.get\_band\_structure()}.

\subsection{E(3)-equivariant message passing, readout heads, and architecture variants}
\label{subsec:architecture}

\begin{figure}[H]
\centering
\includegraphics[width=\textwidth]{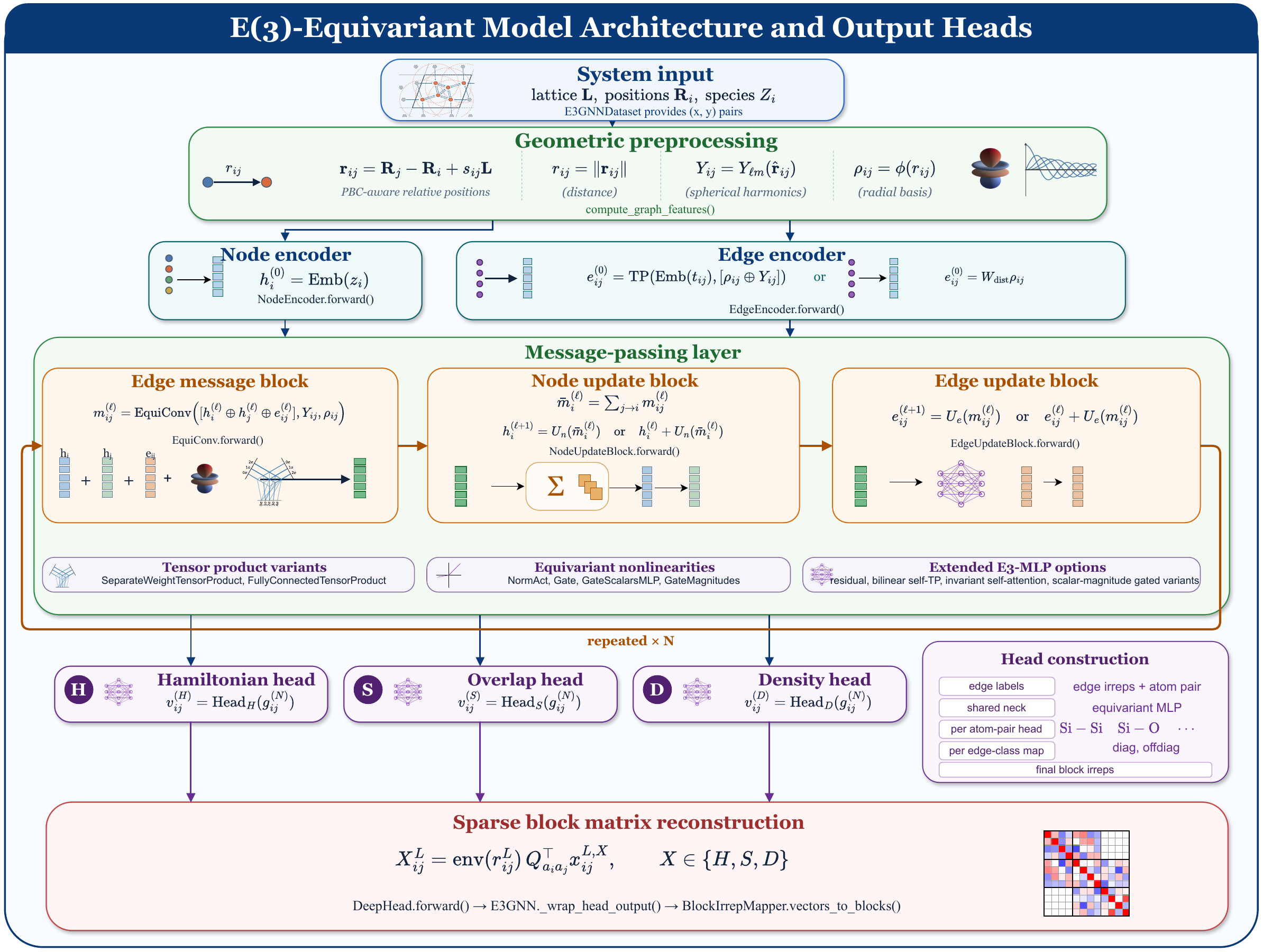}
\caption{E(3)-equivariant model architecture and operator-specific output heads. Periodic relative positions are expanded into radial features and spherical harmonics before node and edge encoders initialize irrep-valued latent states. Repeated message-passing layers form equivariant edge messages, aggregate them into node updates, and update edge features using configurable tensor products, nonlinearities, and residual paths. After the final layer, shared processing can branch by target operator, chemical pair, and edge class. The Hamiltonian, overlap, and density heads produce symmetry-adapted coefficients that the \texttt{BlockIrrepMapper} reconstructs as sparse orbital blocks, with an optional radial envelope applied during readout.}
\label{fig:model_architecture}
\end{figure}

The network is composed of node encoders, edge encoders, a stack of equivariant message-passing blocks, and matrix-specific readout heads.
Figure~\ref{fig:model_architecture} summarizes the computational structure and configuration points of the model.
We first introduce the base architecture and then describe the additional variants that extend it.
If $z_i$ denotes the atomic-species index of atom $i$, the initial node representation is
\begin{equation}
\bm{h}_i^{(0)}
=
\mathrm{Emb}_{\mathrm{atom}}(z_i)
\in d_0 \times 0e,
\end{equation}
so node encoders initialize scalar chemical features. One available edge-encoder variant first forms an edge-type embedding $\bm{u}_{ij}=\mathrm{Emb}_{\mathrm{edge}}(t_{ij})$ and a geometric feature $\bm{g}_{ij}=\varphi(r_{ij}) \oplus Y(\hat{\bm{r}}_{ij})$, and then applies
\begin{equation}
\bm{e}_{ij}^{(0)}
=
\mathrm{TP}_{\mathrm{enc}}\!\left(\bm{u}_{ij},\bm{g}_{ij}\right).
\end{equation}
If $\bm{h}_i^{(t)}$ denotes node features and $\bm{e}_{ij}^{(t)}$ denotes edge features at layer $t$, a generic \texttt{Mandala}
message-passing step may be written as
\begin{align}
\bm{m}_{ij}^{(t)} &=
\Phi_{\mathrm{edge}}^{(t)}
\left(
\bm{h}_i^{(t)}, \bm{h}_j^{(t)}, \bm{e}_{ij}^{(t)},
Y(\hat{\bm{r}}_{ij}), \varphi(r_{ij})
\right), \\
\bm{h}_i^{(t+1)} &=
\Phi_{\mathrm{node}}^{(t)}
\left(
\bm{h}_i^{(t)},
\mathrm{Agg}_{j \in \mathcal{N}(i)} \, \bm{m}_{ij}^{(t)}
\right),
\end{align}
Here, $\bm{h}_i^{(t)}$ denotes the irrep-valued latent feature carried by node $i$ at message-passing layer $t$, $\bm{e}_{ij}^{(t)}$ denotes the corresponding edge feature for the directed edge $i \rightarrow j$, $\bm{m}_{ij}^{(t)}$ is the edge message constructed at that layer, $\hat{\bm{r}}_{ij}$ is the unit vector along the relative displacement between atoms $i$ and $j$, $Y(\hat{\bm{r}}_{ij})$ denotes the spherical-harmonic embedding of that direction, and $\varphi(r_{ij})$ denotes a radial distance embedding.
The neighbor set $\mathcal{N}(i)$ contains the atoms connected to node $i$ within the graph cutoff.
In this base form, $\Phi_{\mathrm{edge}}^{(t)}$ and $\Phi_{\mathrm{node}}^{(t)}$ are equivariant maps assembled from tensor products, equivariant linear layers, radial MLPs, parity-aware nonlinearities, and optional residual connections, while $\mathrm{Agg}$ denotes a configurable aggregation rule such as summation or attention.

In the concrete implementation, these updates are built from equivariant convolutions with radial weighting. Writing the concatenated local feature as
\begin{equation}
\bm{f}_{ij}^{(t)}
=
\bm{h}_i^{(t)} \oplus \bm{h}_j^{(t)} \oplus \bm{e}_{ij}^{(t)},
\end{equation}
the tensor-product stage produces
\begin{equation}
\bm{z}_{ij}^{(t)}
=
\mathrm{TP}\!\left(
\bm{f}_{ij}^{(t)},
Y(\hat{\bm{r}}_{ij})
\right),
\end{equation}
followed by an optional equivariant nonlinearity
\begin{equation}
\widetilde{\bm{z}}_{ij}^{(t)}
=
\mathrm{Act}_{\mathrm{eq}}\!\left(\bm{z}_{ij}^{(t)}\right),
\end{equation}
and a radial MLP that returns one scalar coefficient per output irrep block,
\begin{equation}
\bm{w}_{ij}^{(t)}
=
\mathrm{MLP}_{\mathrm{rad}}\!\left(\varphi(r_{ij})\right).
\end{equation}
If $\widetilde{\bm{z}}_{ij}^{(t,a)}$ denotes the $a$th output irrep slice, the weighted equivariant convolution is
\begin{equation}
\bm{q}_{ij}^{(t,a)}
=
w_{ij}^{(t,a)}\,\widetilde{\bm{z}}_{ij}^{(t,a)}.
\end{equation}
The node update then aggregates edge messages as
\begin{equation}
\overline{\bm{m}}_i^{(t)}
=
\sum_{j\in\mathcal{N}(i)}
\bm{q}_{ji}^{(t)},
\end{equation}
while edge and node updates both admit optional self-connections and residual paths.
In the sequential message block used by the code, node features are first updated from the current edges and only then are edge features updated using the new node state.

The readout stage maps final latent features to the irrep content required by each matrix target.
For a target operator $X$, the head predicts irrep vectors
\begin{equation}
\widehat{\bm{x}}_{ij}^{\bm{L},X}
=
\Psi_X\!\left(
\bm{e}_{ij}^{(L)},
\bm{h}_i^{(L)},
\bm{h}_j^{(L)},
\varphi(r_{ij})
\right),
\end{equation}
followed by inverse irrep-to-block conversion
\begin{equation}
X_{ij}^{\bm{L},\mathrm{pred}}
=
Q_{\mathcal{B}_i\mathcal{B}_j}^{\top}\widehat{\bm{x}}_{ij}^{\bm{L},X},
\end{equation}
where $\mathcal{B}_i$ and $\mathcal{B}_j$ denote the complete species-specific orbital and shell metadata. This distinction permits different basis variants for the same element in different contexts.
For richer readout variants, the edge used by the head may first be transformed through a tensor-square map,
\begin{equation}
\bm{t}_{ij}^{(L)}
=
\mathrm{TS}\!\left(\bm{e}_{ij}^{(L)}\right),
\end{equation}
with the actual head input defined by
\begin{equation}
\bm{u}_{ij}^{(L)}
=
\begin{cases}
\bm{t}_{ij}^{(L)}, & \text{tensor-square head enabled},\\
\bm{e}_{ij}^{(L)}, & \text{otherwise}.
\end{cases}
\end{equation}
The head can then branch by edge class,
\begin{equation}
\widehat{\bm{x}}_{ij}^{\bm{L},X}
=
\begin{cases}
\Psi^{X}_{\mathrm{diag}}(\bm{u}_{ij}^{(L)}), &
(i,j,\bm{L})\in\mathcal{E}_{\mathrm{diag}}, \\
\Psi^{X}_{\mathrm{shifted\_self}}(\bm{u}_{ij}^{(L)}), &
(i,j,\bm{L})\in\mathcal{E}_{\mathrm{shifted\_self}}, \\
\Psi^{X}_{\mathrm{offdiag}}(\bm{u}_{ij}^{(L)}), &
(i,j,\bm{L})\in\mathcal{E}_{\mathrm{offdiag}}.
\end{cases}
\end{equation}
The most elaborate readout variants further split scalar and non-scalar output channels or factorize each predicted block into a normalized equivariant direction and a positive scalar magnitude. In the latter case one predicts
\begin{equation}
\widehat{\bm{x}}_{e,\mathrm{norm}}^{X}
=
\Psi_{\mathrm{dir}}^{X}(\bm{u}_{e}),
\qquad
\widehat{m}_{e}^{X}
=
\exp\!\left(
\mathrm{MLP}_{\mathrm{mag}}^{X}
\bigl(
L_{\mathrm{mag}}^{X}(\bm{t}_{e})
\oplus
\varphi(r_e)
\bigr)
\right),
\end{equation}
and reconstructs the block prediction as
\begin{equation}
\bm{x}_{e}^{X,\mathrm{pred}}
=
\widehat{m}_{e}^{X}\,
\widehat{\bm{x}}_{e,\mathrm{norm}}^{X}.
\end{equation}
Because heads share the same equivariant latent representation, \texttt{Mandala} can predict Hamiltonian, overlap, and density matrices jointly with only modest additional readout cost.

This canonical architecture is the reference point for the configurable variants summarized in Table~\ref{tab:architecture-variants}. All variants retain the same sparse data model, block-to-irrep mapping, and training interface, so a controlled comparison changes the selected component without changing the surrounding workflow. Configurations that reproduce the DeepH, DeepH-E3, and MACE-H architectures, together with their inference efficiency, are described in \mbox{\ref{app:reference-architectures}}.

\begin{table}[H]
\centering
\small
\caption{Compact architecture-variant surface exposed by the common \texttt{Mandala} model.}
\label{tab:architecture-variants}
\begin{tabular}{p{0.27\textwidth}p{0.64\textwidth}}
\toprule
Component & Supported choices \\
\midrule
Latent representation & Hidden irrep multiplicities, $\ell_{\max}$, radial basis size, and message-passing depth \\
Edge construction & Rich tensor-product or distance-linear encoder; optional spherical-harmonic tensor square \\
Equivariant convolution & Separate-weight or fully connected tensor products; configurable equivariant nonlinearities \\
Node aggregation & Sum, mean, or invariant attention \\
Update paths & Direct or residual node and edge updates; optional self-connections \\
Readout & Shared neck with operator-, atom-pair-, and edge-class branches; optional tensor-square or magnitude factorization \\
Radial behavior & Direct prediction or multiplication by a fitted pair-dependent envelope \\
\bottomrule
\end{tabular}
\end{table}

\subsection{Sparse operator algebra, symmetrization, gauge alignment, and density normalization}
\label{subsec:postprocessing}

Sparse operator manipulations are implemented directly on aligned block data.
Given two sparse operators $A$ and $B$ with reverse-edge alignment already established, \texttt{Mandala} evaluates their trace contraction as
\begin{equation}
\Tr(AB)
=
\sum_{(\bm{L},i,j)}
\mathrm{tr}\!\left(A_{ij}^{\bm{L}} B_{ji}^{-\bm{L}}\right),
\end{equation}
without assembling dense global matrices. This operator algebra is reused for energy and electron-count evaluation, loss construction, and diagnostic analysis.
For efficient vectorized evaluation, the implementation precomputes a reverse-edge alignment permutation. For each pair key $k$ with reverse key
$k^{\mathrm{rev}}$, the permutation $P_k$ is defined by
\begin{equation}
P_k(n)=m
\quad\Longleftrightarrow\quad
\bigl(\bm{L}_{k,n},i_{k,n},j_{k,n}\bigr)
=
\bigl(-\bm{L}_{k^{\mathrm{rev}},m},j_{k^{\mathrm{rev}},m},i_{k^{\mathrm{rev}},m}\bigr),
\end{equation}
so that the aligned trace may be written as
\begin{equation}
\Tr(AB)
=
\sum_k \sum_n
\mathrm{tr}\!\left(
A_{k,n}\,
B_{k^{\mathrm{rev}},P_k(n)}
\right).
\end{equation}

For matrix targets expected to be symmetric or Hermitian in the chosen basis, \texttt{Mandala} supports explicit sparse symmetrization after block reconstruction.
In the real-valued convention used by the implementation, this takes the form
\begin{equation}
X_{ij}^{\bm{L}}
\leftarrow
\frac{1}{2}
\left(
X_{ij}^{\bm{L}} + \left(X_{ji}^{-\bm{L}}\right)^{\top}
\right).
\end{equation}
This operation is carried out using precomputed reverse-edge permutations, so that symmetry is enforced consistently across periodic images and self-edges.

Hamiltonian gauge freedom is handled during reference-based evaluation through the overlap-shift alignment described in Section~\ref{subsubsec:gauge_invariance}. For matrix-error reporting, \texttt{Mandala} computes
\begin{equation}
\widetilde{H}^{\mathrm{pred}}
=
H^{\mathrm{pred}} - \mu_H S,
\qquad
\mu_H
=
\frac{
\left\langle H^{\mathrm{pred}} - H^{\mathrm{ref}}, S \right\rangle_F
}{
\left\langle S, S \right\rangle_F
},
\end{equation}
so that purely gauge-like Hamiltonian offsets are removed before reporting gauge-aligned operator errors. This expression is not used for target-free inference because it requires the reference Hamiltonian.

The framework also supports density normalization to the correct number of electrons.
If a predicted density $D^{\mathrm{pred}}$ and overlap $S^{\mathrm{pred}}$ produce $N_e^{\mathrm{pred}} = \Tr(D^{\mathrm{pred}}S^{\mathrm{pred}})$, then a normalized density is obtained by the scalar rescaling
\begin{equation}
D_{\mathrm{norm}}^{\mathrm{pred}}
=
\beta D^{\mathrm{pred}},
\qquad
\beta = \frac{N_e^{\mathrm{target}}}{\Tr(D^{\mathrm{pred}}S^{\mathrm{pred}})}.
\end{equation}
This provides a simple and effective way to enforce electron-count consistency.

For generalized eigensolves, the overlap can additionally be conditioned by Hermitian symmetrization followed by an eigenvalue projection. If $S=U\operatorname{diag}(s_i)U^\dagger$, the conditioned matrix is
\begin{equation}
S_{\mathrm{PSD}}
=U\operatorname{diag}\!\left(\max(s_i,\varepsilon_S)\right)U^\dagger,
\qquad \varepsilon_S=10^{-6}.
\end{equation}
Thus negative and near-zero eigenvalue components are removed from the admissible spectrum, with a small positive floor retained for numerical factorization. This safeguard is optional and its activation is recorded in spectral outputs. A Cholesky failure is reported if neither this projection nor the separately configurable diagonal-jitter fallback produces a valid factorization.

These operations enforce aligned transpose symmetry, an optional positive-definite numerical overlap for eigensolution, and a scalar electron-count normalization. They do not by themselves enforce density-matrix positivity, occupation bounds, generalized idempotency, or mutual consistency among independently predicted $H$, $S$, and $D$. The resulting symmetrized and normalized operators are suitable for the downstream diagnostics described below, while these remaining constraints define limitations of the present model.
Their place in the analysis pipeline is summarized in Figure~\ref{fig:postprocessing_analysis} after the training objective is introduced.

\subsection{Observable guidance with band energies and electron counts}
\label{subsec:training_objectives}

\begin{figure}[H]
\centering
\includegraphics[width=\textwidth]{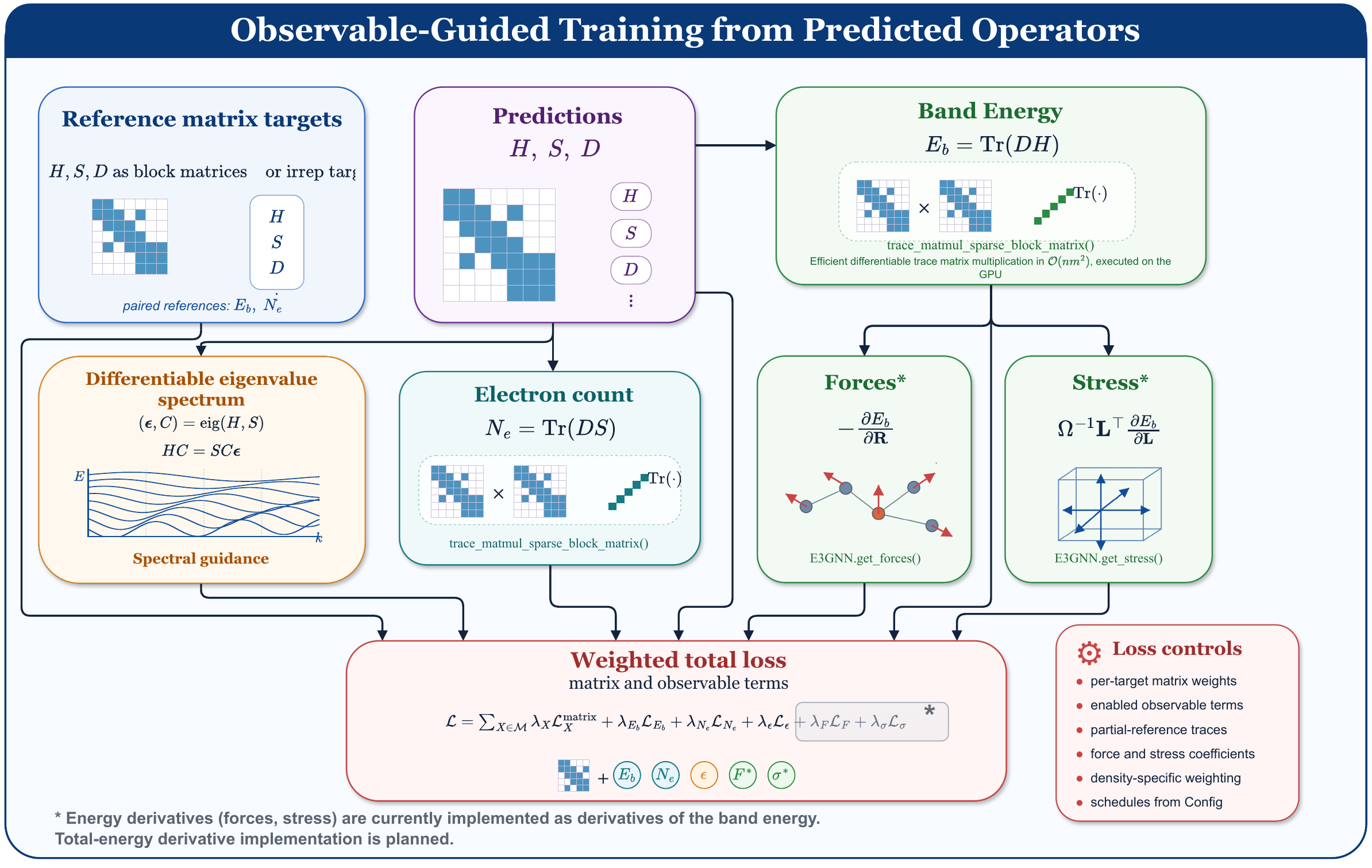}
\caption{Observable-guided training from predicted sparse operators. Hamiltonian, overlap, and density predictions may be supervised directly as blocks or irrep coefficients. Because the operators remain differentiable, the same predictions also define band energy, electron count, and generalized-eigenvalue spectra, allowing enabled observable errors to enter the weighted training objective without auxiliary prediction heads. Configuration controls the active terms, their weights, and their schedules. The available force and stress pathways differentiate the band energy only; they do not yet represent derivatives of the Kohn--Sham total energy.}
\label{fig:observable_guidance}
\end{figure}

\texttt{Mandala} supports pure matrix learning, mixed matrix-and-observable learning, and fully observable-guided training from one common operator-prediction pipeline.
Figure~\ref{fig:observable_guidance} summarizes the corresponding training flow.
For a chosen set of matrix targets $\mathcal{M} \subseteq \{H,S,D\}$, the general training objective is
\begin{equation}
\Loss
=
\sum_{X \in \mathcal{M}} \lambda_X \Loss_X^{\mathrm{matrix}}
+ \lambda_E \Loss^{E}
+ \lambda_N \Loss^{N}
+ \lambda_{\mathrm{spec}}\Loss^{\mathrm{spec}}.
\end{equation}
The matrix terms compare predicted and reference sparse operators, typically by a weighted combination of $\ell_1$ and $\ell_2$ errors,
\begin{equation}
\Loss_X^{\mathrm{matrix}}
=
\eta \, \mathrm{MAE}(X^{\mathrm{pred}},X^{\mathrm{ref}})
+ (1-\eta)\, \mathrm{MSE}(X^{\mathrm{pred}},X^{\mathrm{ref}}),
\end{equation}
with $\eta$ controlled by configuration (usually $1.0$ or $0.0$ to select MAE or MSE).

Observable guidance is defined on quantities derived from the predicted operators.
In the operator-centered workflow used throughout \texttt{Mandala}, the band energy and the number of electrons are
\begin{equation}
E_b^{\mathrm{pred}} = \Tr(D^{\mathrm{pred}}H^{\mathrm{pred}}),
\qquad
N_e^{\mathrm{pred}} = \Tr(D^{\mathrm{pred}}S^{\mathrm{pred}}),
\end{equation}
which gives
\begin{equation}
\Loss^E = \mathrm{MSE}(E_b^{\mathrm{pred}},E_b^{\mathrm{ref}}),
\qquad
\Loss^N = \mathrm{MSE}(N_e^{\mathrm{pred}},N_e^{\mathrm{ref}}).
\end{equation}
The framework also supports mixed observable guidance paths in which one of the operators entering the trace is replaced by the reference operator, for example the diagnostic contractions
\begin{equation}
E_{H|D^{\mathrm{ref}}}^{\mathrm{pred}}
=
\Tr\!\left(H^{\mathrm{pred}} D^{\mathrm{ref}}\right),
\qquad
E_{D|H^{\mathrm{ref}}}^{\mathrm{pred}}
=
\Tr\!\left(H^{\mathrm{ref}} D^{\mathrm{pred}}\right),
\end{equation}
and
\begin{equation}
N_{D|S^{\mathrm{ref}}}^{\mathrm{pred}}
=
\Tr\!\left(D^{\mathrm{pred}} S^{\mathrm{ref}}\right),
\qquad
N_{S|D^{\mathrm{ref}}}^{\mathrm{pred}}
=
\Tr\!\left(D^{\mathrm{ref}} S^{\mathrm{pred}}\right),
\end{equation}
enabling diagnostic and training modes that separate errors due to different predicted operators.

The current implementation also exposes experimental derivatives of the operator-derived band energy with respect to atomic coordinates and the cell,
\begin{equation}
\bm{F}^{\mathrm{band}}_I
=
-\frac{\partial E_b^{\mathrm{pred}}}{\partial \bm{R}_I},
\qquad
\sigma^{\mathrm{band}}
=
\frac{1}{\Omega}h^{\top}
\frac{\partial E_b^{\mathrm{pred}}}{\partial h},
\qquad \Omega=\det(h).
\end{equation}
These derivatives are not total-energy forces or stresses because $E_b^{\mathrm{pred}}=\Tr(D^{\mathrm{pred}}H^{\mathrm{pred}})$ omits the remaining contributions to the Kohn--Sham total energy.
They are therefore not presented here as validated supervision targets.

The supported observable-guidance objective in this work is thus restricted to matrix entries, the band energy, and the electron count.
Once the remaining total-energy contributions are implemented, the existing automatic-differentiation pathway can be extended to total-energy force and stress supervision without changing the sparse-operator representation.

\subsection{Spectral guidance}
\label{subsec:spectral-guidance}

For periodic systems, the spectral term compares eigenvalues derived from the predicted real-space Hamiltonian with reference eigenvalues on a uniform Monkhorst--Pack-style mesh. For mesh dimensions $N_1\times N_2\times N_3$, the implementation uses fractional points
\begin{equation}
\bm{q}_{n_1n_2n_3}
=
\left(\frac{n_1}{N_1},\frac{n_2}{N_2},\frac{n_3}{N_3}\right),
\qquad 0\le n_\alpha<N_\alpha,
\end{equation}
and assembles the reciprocal-space operators from the shift-resolved blocks,
\begin{equation}
H(\bm{q})=\sum_{\bm{L}}e^{i\bm{k}(\bm{q})\cdot\bm{R}_{\bm{L}}}H^{\bm{L}},
\qquad
S(\bm{q})=\sum_{\bm{L}}e^{i\bm{k}(\bm{q})\cdot\bm{R}_{\bm{L}}}S^{\bm{L}}.
\end{equation}
The spectral loss uses the reference overlap for both spectra. This isolates the Hamiltonian contribution and avoids conflating spectral guidance with overlap error. After Hermitian symmetrization, $S(\bm{q})=LL^\dagger$ is factorized by Cholesky decomposition and the generalized eigenvalues are obtained from
\begin{equation}
\epsilon(\bm{q})
=
\operatorname{eigvalsh}\!\left(
L^{-1}H(\bm{q})L^{-\dagger}
\right).
\end{equation}
Eigenvalues are sorted in ascending order by the Hermitian solver and matched by band index; no separate band-tracking heuristic is applied.

Only bands near the reference Fermi level $E_F^{\mathrm{ref}}$ contribute. With a central half-width $W$ and cosine taper $\tau$, the band weight is one for
$|\epsilon^{\mathrm{ref}}-E_F^{\mathrm{ref}}|\le W-\tau$, decreases smoothly to zero over the outer interval of width $\tau$, and vanishes outside $W$. Uniform k-point weights are used. The implemented objective is therefore
\begin{equation}
\Loss^{\mathrm{spec}}
=
\frac{\sum_{\bm{q},b}w_{\bm{q}}w_{\bm{q}b}
\rho_\delta\!\left(
\epsilon^{\mathrm{pred}}_{\bm{q}b}
-\epsilon^{\mathrm{ref}}_{\bm{q}b}
\right)}
{\sum_{\bm{q},b}w_{\bm{q}}w_{\bm{q}b}},
\end{equation}
where $\rho_\delta$ is configurable as Huber, absolute, or squared error. The ablation in Section~\ref{sec:ablations} uses a $2\times2\times2$ mesh, $W=10$~eV, $\tau=2$~eV, and Huber loss with $\delta=0.1$~eV. Both spectra are expressed relative to the same reference Fermi level; this changes the displayed reference but not their difference and does not introduce a reference-dependent fitted gauge shift.

The Fourier assembly, Cholesky transformation, and Hermitian eigensolve remain in the automatic-differentiation graph, so gradients propagate to the predicted real-space blocks. Reference factorizations and spectra may be cached.

\begin{figure}[H]
\centering
\includegraphics[width=\textwidth]{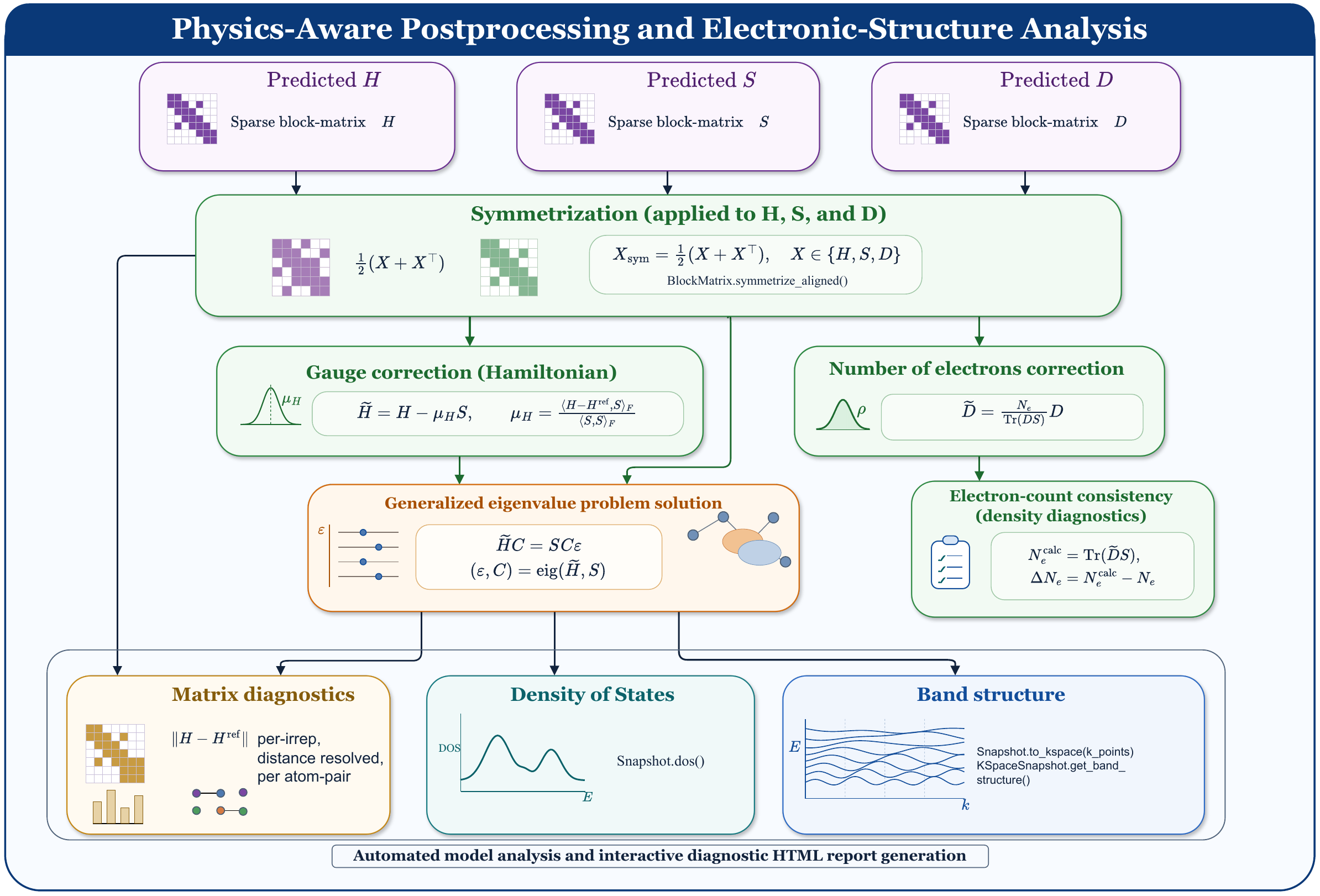}
\caption{Physics-aware postprocessing and electronic-structure analysis. Predicted Hamiltonian, overlap, and density blocks are symmetrized in their aligned sparse representation. For comparison with a reference, the Hamiltonian may be aligned along the overlap-proportional energy gauge; the density may be rescaled to reproduce the prescribed electron count. The Hamiltonian and overlap enter the generalized eigenvalue problem used for density-of-states and band-structure calculations, with optional overlap conditioning, while the density is checked through its overlap-weighted trace. Matrix, irrep, chemical-pair, and distance-resolved diagnostics, together with the spectral analyses, are collected in the automated model report.}
\label{fig:postprocessing_analysis}
\end{figure}

After prediction, the operators pass through the physical corrections and analysis routes shown in Figure~\ref{fig:postprocessing_analysis}.
Symmetrization acts on all predicted matrices, while reference-based Hamiltonian gauge alignment and density normalization address operator-specific evaluation and normalization steps.
The corrected Hamiltonian and overlap determine spectral quantities through the generalized eigenvalue problem, whereas the normalized density provides an independent electron-count consistency check.
The same processed objects feed both scalar summaries and resolved diagnostics, allowing automated reports to connect aggregate errors with their chemical, spatial, and spectral origins.

\subsection{Configuration-driven architecture search}
\label{subsec:search}

Because the same sparse operator workflow supports many architectural choices, \texttt{Mandala} treats model search as a primary scientific use case rather than as an auxiliary engineering convenience.
All major representation, encoder, message-passing, readout, and loss-weighting choices are exposed as structured configuration parameters and can therefore be swept systematically.
We use Weights \& Biases (W\&B) to launch and track sweeps, retain run configurations and checkpoint references, and compare completed trials~\cite{wandb}.

Search dimensions include hidden irreps, angular cutoff, radial basis size, message-passing depth, encoder style, tensor-product type, normalization and nonlinearity families, node and edge update variants, residual pathways, neck and head depth, separate treatment of shifted-self blocks, tensor-square readout options, scalar-specific branches, observable-loss coefficients, and the subset of matrix targets predicted jointly.
This wide configuration surface is one of the main scientific strengths of the framework:
it enables controlled comparisons between architecture families without forcing changes to the data model, parser layer, or observable-evaluation stack.
The importance of systematic hyperparameter exploration in machine-learned electronic-structure models has also been emphasized in related work on training-free optimization strategies \cite{hyperparameterpaper}.

\subsection{Computational complexity and inference scaling}
\label{subsec:performance}

For a fixed real-space cutoff, the number of graph edges $|\mathcal{E}|$ grows linearly with the number of atoms $N$ at approximately constant density. Sparse graph preparation and storage therefore scale as $\mathcal{O}(N+|\mathcal{E}|)$. For a fixed irrep layout and number of layers, message passing scales as $\mathcal{O}(|\mathcal{E}|C_{\mathrm{TP}})$, where $C_{\mathrm{TP}}$ is the cost of the configured tensor-product paths. Readout and sparse block reconstruction scale with the number and dimensions of the retained atom-pair blocks. Dense downstream diagonalization is separate from network inference and scales cubically with the number of basis functions; it is not included in the benchmark below.

\begin{figure}[H]
\centering
\begin{minipage}[t]{0.49\textwidth}
\centering
\includegraphics[width=\linewidth]{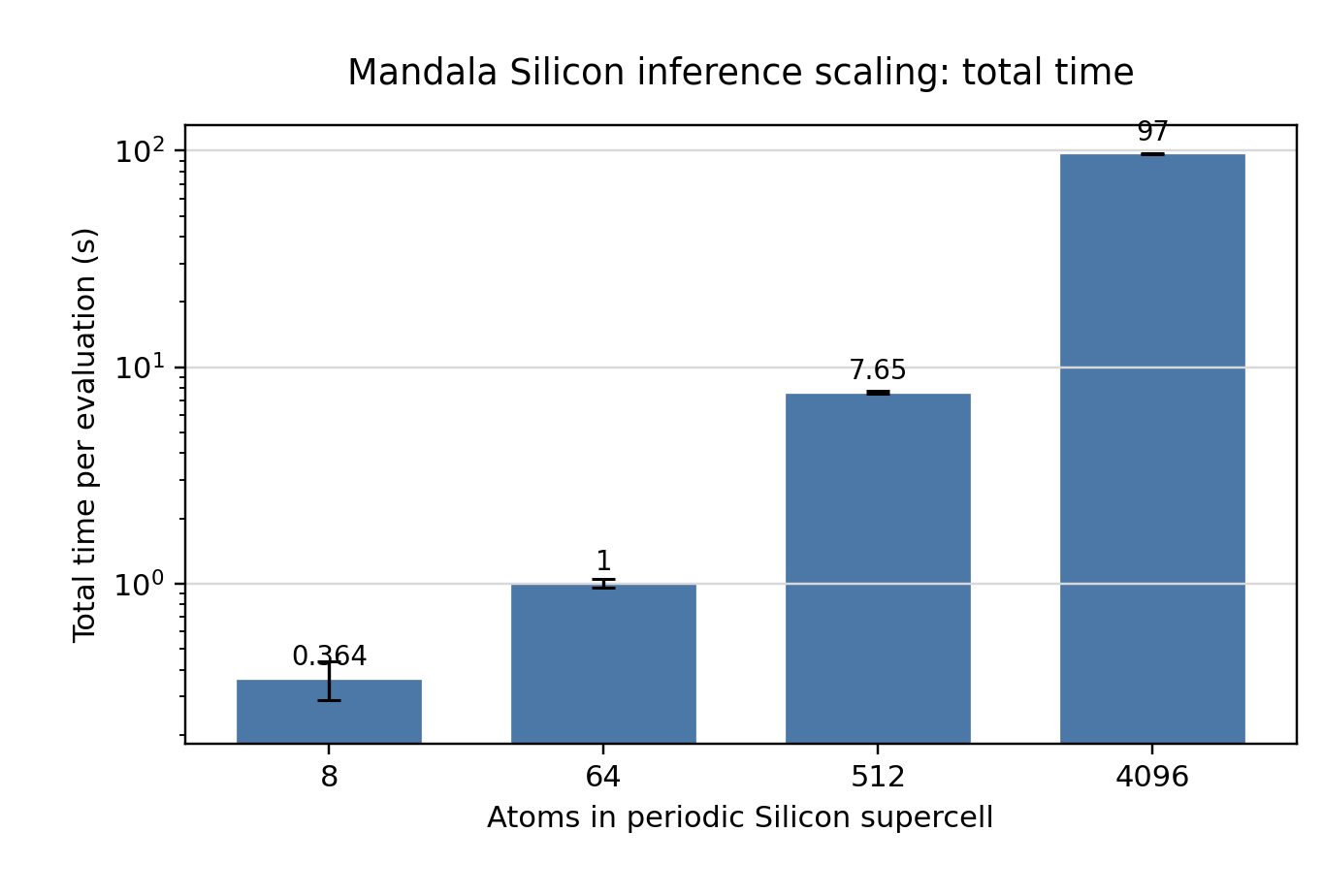}
\small (a) Total wall time
\end{minipage}\hfill
\begin{minipage}[t]{0.49\textwidth}
\centering
\includegraphics[width=\linewidth]{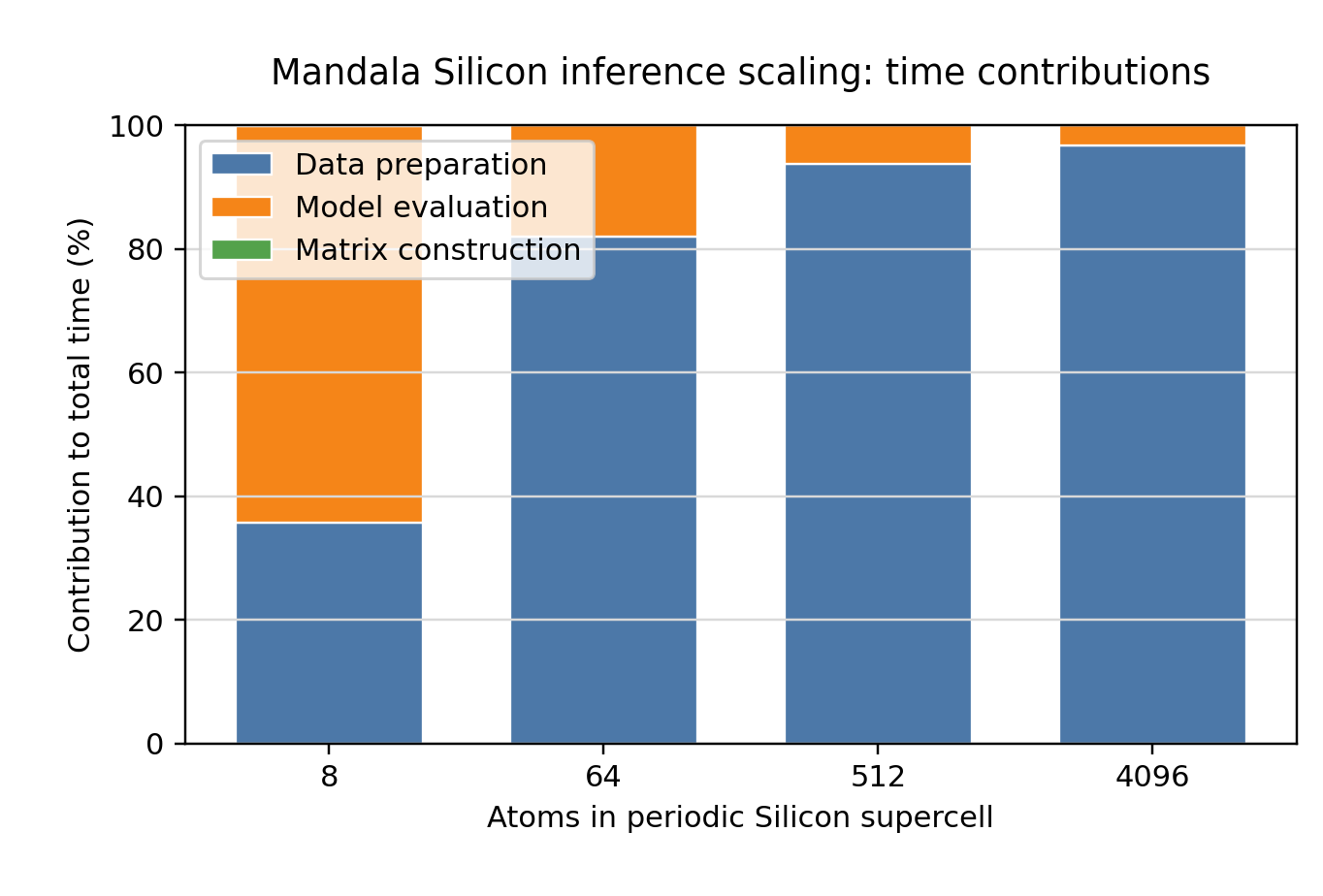}
\small (b) Time contributions
\end{minipage}

\vspace{0.6em}
\begin{minipage}[t]{0.49\textwidth}
\centering
\includegraphics[width=\linewidth]{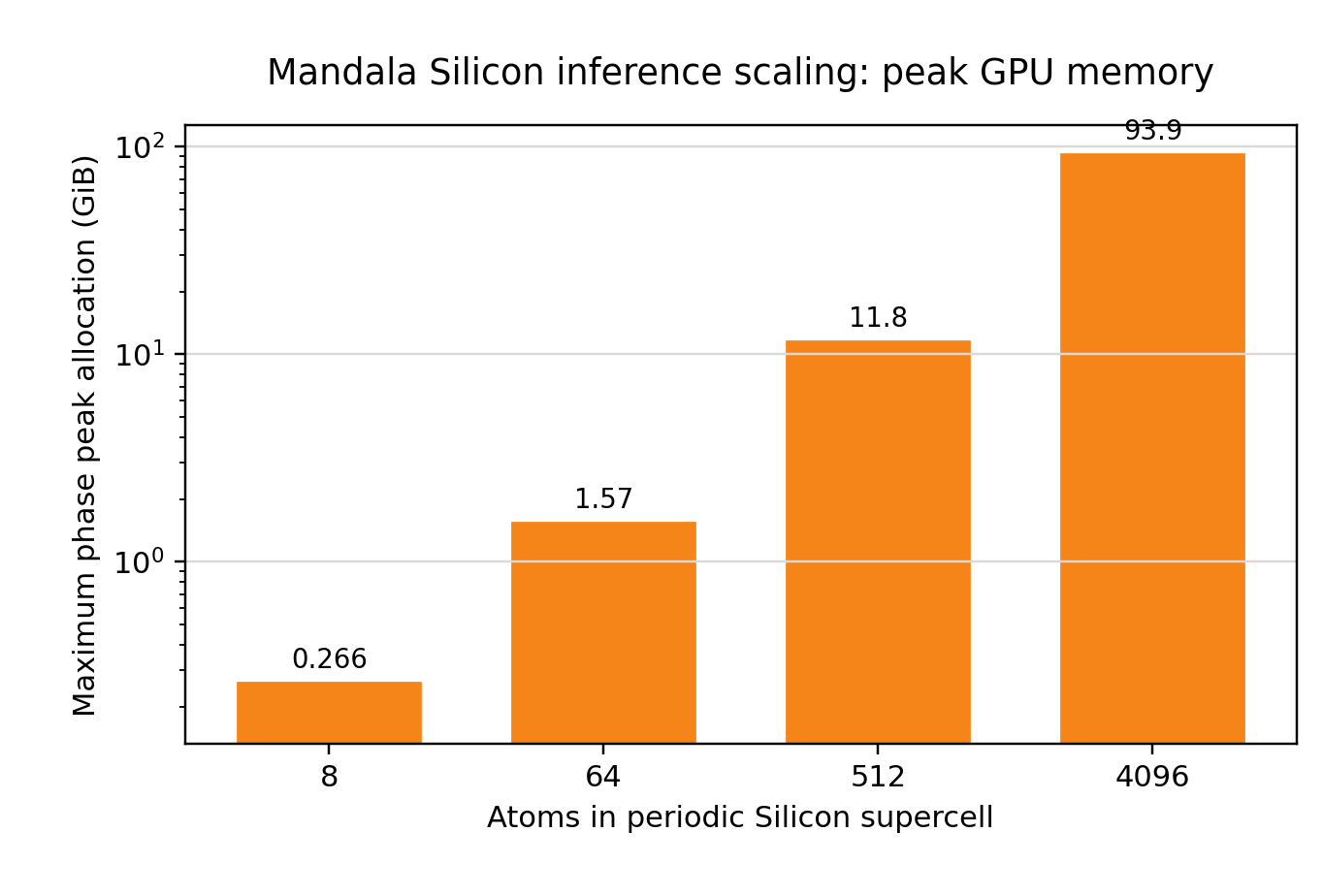}
\small (c) Maximum peak GPU memory
\end{minipage}\hfill
\begin{minipage}[t]{0.49\textwidth}
\centering
\includegraphics[width=\linewidth]{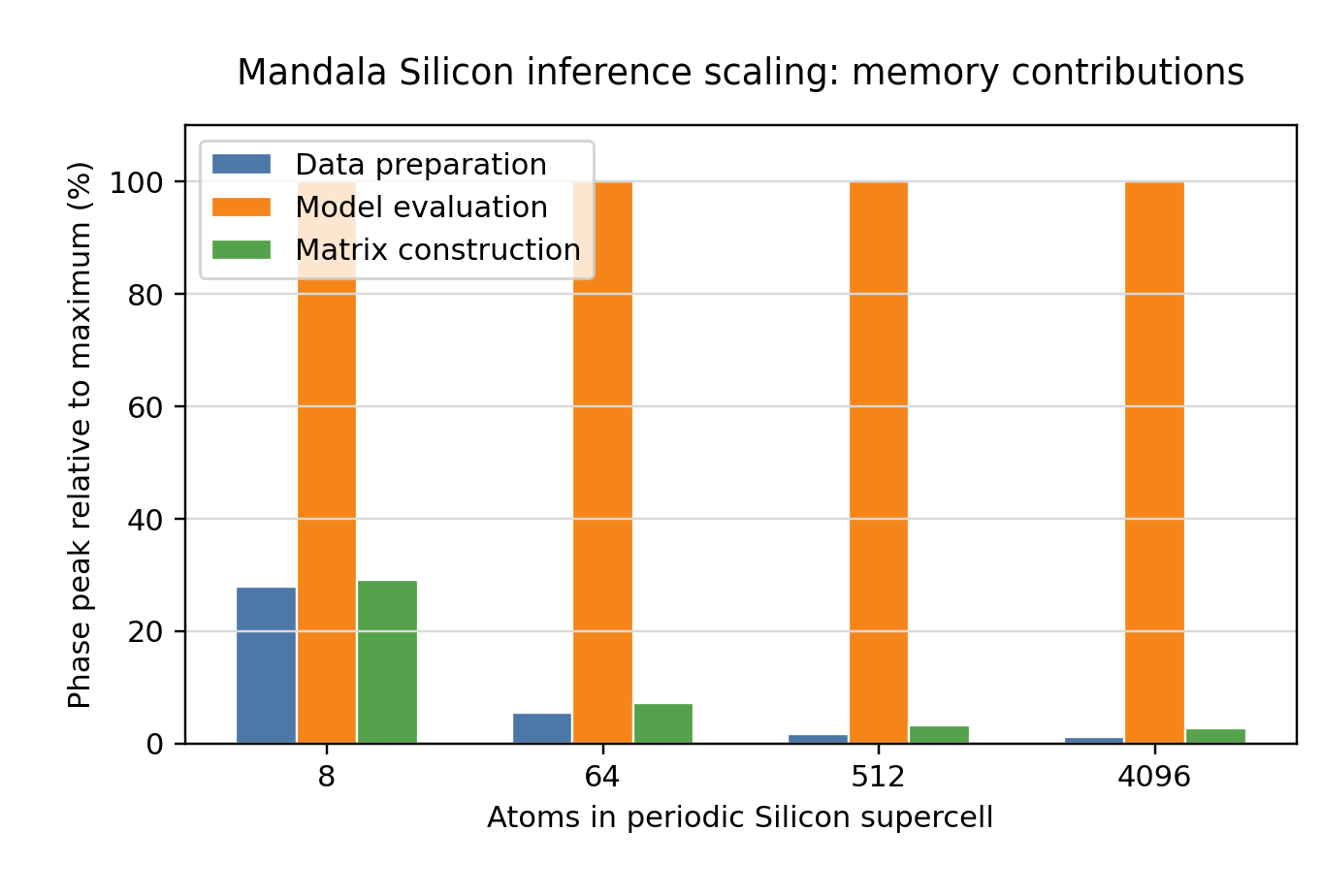}
\small (d) Memory contributions
\end{minipage}
\caption{Inference scaling for periodic silicon supercells on one NVIDIA B200 GPU. Panels (a) and (c) use logarithmic axes for the total time and the largest independently recorded phase-memory peak, respectively. Panels (b) and (d) use linear percentage axes: timing phases sum to the total, whereas each independently measured memory peak is normalized to the largest phase peak at that system size. Timing values are means of ten calls after two warm-ups; error bars in panel (a) show the combined one-standard-deviation variation.}
\label{fig:silicon-scaling}
\end{figure}

Figure~\ref{fig:silicon-scaling} reports target-free inference for periodic diamond-silicon supercells containing 8, 64, 512, and 4096 atoms. The benchmark uses the selected joint Hamiltonian--density--overlap model in single-precision arithmetic on one NVIDIA B200 GPU, an 8~\r{A} cutoff, two warm-up calls, and ten timed repetitions per size. It measures three phases independently: construction of graph features from a CIF structure, the neural-network forward pass, and reconstruction of sparse operator blocks. The corresponding block counts are 2376, 19\,008, 152\,064, and 1\,216\,512 respectively. Absolute totals and maxima are separated from phase composition so that logarithmic scaling does not visually distort the relative contributions.

Total evaluation time increases from $0.364$~s for 8 atoms to $97.0$~s for 4096 atoms. The phase decomposition shows a crossover from model-dominated execution at 8 atoms to data-preparation-dominated execution: graph preparation grows from $0.130$~s to $93.8$~s and accounts for approximately $97\%$ of the largest-system total, while the forward pass grows from $0.234$~s to $3.17$~s. The model-evaluation phase sets the maximum memory requirement at every measured size, increasing from $0.266$~GiB to $93.9$~GiB; the other independently measured phase peaks become progressively smaller fractions of that maximum. These measurements demonstrate sparse model execution to 4096 atoms on the stated hardware and identify input preparation and model-evaluation memory as the present large-system bottlenecks; they are not a training-throughput benchmark.

\subsection{Current limitations and development roadmap}
\label{subsec:limitations}

The present implementation is designed for localized-orbital electronic-structure data that can be represented by real-valued, atom-pair-resolved sparse blocks.
Orbital definitions must be known for every chemical species.
Locality is controlled through a finite cutoff radius; consequently, accuracy and computational cost depend on whether the selected cutoff captures the relevant operator range.
Periodic spectral analysis further requires consistent lattice-shift support and a numerically well-conditioned overlap matrix.

The current training workflow is optimized for a batch size of one, as is common for variable-size operator targets. Larger batches and distributed data parallelism are not yet implemented.
The configuration surface is intentionally broad, but not every combination is equally mature; the controlled ablations reported in this work are therefore important for identifying supported reference settings.
Most importantly, the scalar $\Tr(DH)$ is a band-energy observable rather than the complete Kohn--Sham total energy.
Its coordinate and cell derivatives cannot be interpreted as total-energy forces and stresses.
Adding all total-energy contributions and validating their derivatives is a planned extension.

\subsection{Verification, reproducibility, and extensibility}
\label{subsec:verification}

\texttt{Mandala} includes tests and validation routines covering data parsing, basis harmonization, sparse operator alignment, block-to-irrep mapping, graph construction, equivariant tensor-product paths, observable evaluation, and end-to-end training workflows.
These checks are especially important in a code base where subtle mistakes in basis ordering, reverse-edge alignment, or irrep bookkeeping can produce apparently plausible but physically inconsistent results.

The implementation supports reproducible runs through serialized configurations, deterministic sparse edge ordering, stored splits, and checkpoints; exact numerical reproducibility additionally depends on software versions, hardware, and deterministic backend settings.
Extensibility follows from the layered structure described above: new electronic structure backends, new observable definitions, new equivariant blocks, and new training objectives can be incorporated without changing the core operator data model.
This design allows \texttt{Mandala} to function both as a reusable scientific software package and as a framework for rapid methodological experimentation.

\section{Results}
\label{sec:results}

We demonstrate the framework's capabilities on three datasets chosen to probe multicomponent chemistry, amorphous environments, and joint operator and spectral prediction. These are capability demonstrations rather than a proper benchmark against other methods: one selected \texttt{Mandala} model is shown for each material. The numerical tables report averages over the complete validation split of that model.

\subsection{Demonstration data and evaluation protocol}
\label{subsec:demonstration-protocol}

Table~\ref{tab:dataset-protocol} summarizes the populations used for the three demonstrations. The ZnCu$_2$Sn(SeS)$_2$ corpus contains 100 perturbed 32-atom cells. The silicon corpus contains 100 independently perturbed eight-atom diamond cells. The SiO$_2$ corpus contains 165 MD-produced 150-atom amorphous glass configurations, of which 100 were selected by the deterministic split procedure.

\begin{table}[H]
\centering
\small
\caption{Dataset populations and splits for the selected demonstration models.}
\label{tab:dataset-protocol}
\begin{tabular}{p{0.28\textwidth}p{0.12\textwidth}p{0.16\textwidth}cp{0.19\textwidth}}
\toprule
Dataset & Corpus & Used split & Seed & Example \\
\midrule
ZnCu$_2$Sn(SeS)$_2$ & $100$ & 80/20 & 44 & no.~94\\
Amorphous SiO$_2$ & 165 & 90/10 & 42 & 13\_67164000 \\
Perturbed Si & $100$ & 80/20 & 43 & no.~90 \\
\bottomrule
\end{tabular}
\end{table}

All reference operators were calculated with OpenMX~3.9 using the PBE exchange--correlation functional, norm-conserving PBE19 pseudopotentials, scalar non-spin-polarized calculations. Table~\ref{tab:reference-settings} gives the material-specific basis and Brillouin-zone settings. Self-consistent calculations used a $10^{-8}$~Hartree convergence threshold. The sparse targets were converted to the common e3nn real spherical-harmonic convention and filtered with the same real-space cutoff used to build the model graph. The checkpoint with the lowest validation Hamiltonian MAE was retained. Each displayed model is one training run; seed replication is used for the controlled ablations rather than for the showcase models.

\begin{table}[H]
\centering
\scriptsize
\caption{Reference electronic-structure and model-support settings. Basis strings use the OpenMX notation.}
\label{tab:reference-settings}
\begin{tabular}{p{0.19\textwidth}p{0.70\textwidth}}
\toprule
Dataset & Reference calculation and model support \\
\midrule
ZnCu$_2$Sn(SeS)$_2$ &
Zn10.0S-s2p2d2/Zn\_PBE19S; Cu10.0S-s2p2d2/Cu\_PBE19S; Sn7.0-s2p2d3f1/Sn\_PBE19; Se9.0-s4p3d3f2/Se\_PBE19; S9.0-s4p3d3f2/S\_PBE19; 400~Ry grid; $11^3$ k mesh; 11~\r{A} model cutoff \\
SiO$_2$ &
Si9.0-s2p2d1/Si\_PBE19; O7.0-s2p2d1/O\_PBE19; 600~Ry grid; $\Gamma$-point sampling; 10~\r{A} model cutoff \\
Perturbed Si &
Si7.0-s2p2d1/Si\_PBE19; 400~Ry grid; $8^3$ k mesh; 8~\r{A} model cutoff \\
\bottomrule
\end{tabular}
\end{table}

Hamiltonian MAEs are shift-resolved sparse-block errors after transpose symmetrization and are reported in electronvolts. Density and overlap errors use the same matrix-element definition in their native dimensionless representations. The band-energy diagnostic for the Hamiltonian-only SiO$_2$ model contracts $H^{\mathrm{pred}}$ with $D^{\mathrm{ref}}$. The spectral figures use the reference overlap matrix, which is inexpensive to obtain from the underlying localized-basis calculation. Energies are shown relative to the reference Fermi level.

Each relative value in the following tables uses the same structures as its absolute validation error. Specifically, it is the mean per-structure validation MAE divided by the mean absolute reference value over that complete validation split. For matrix targets the denominator is the corresponding mean absolute matrix-element magnitude; for band energy it is the mean absolute reference band energy per atom. Further error statistics for all three models, including percentiles and maximum errors, are given in \mbox{\ref{app:error-distributions}}.

\subsection{Multicomponent ZnCu$_2$Sn(SeS)$_2$}

ZnCu$_2$Sn(SeS)$_2$ combines five elements and chemically distinct cation and anion environments in one periodic material.
Its electronic structure contains many element-pair channels with different orbital content, so a single model must resolve both the heterogeneous local chemistry and the coupling between those environments.
The aggregate validation result is summarized in Table~\ref{tab:zncusnses-validation}; the compact configuration of the evaluated model is provided in \ref{app:model-config-zncusnses}.

\begin{table}[H]
\centering
\caption{Validation metrics for the ZnCu$_2$Sn(SeS)$_2$ model.}
\label{tab:zncusnses-validation}
\begin{tabular}{lc}
\toprule
Metric & Value (relative) \\
\midrule
Hamiltonian MAE & 0.003012~eV (2.95\%) \\
\bottomrule
\end{tabular}
\end{table}

\begin{figure}[H]
\centering
\includegraphics[width=\textwidth]{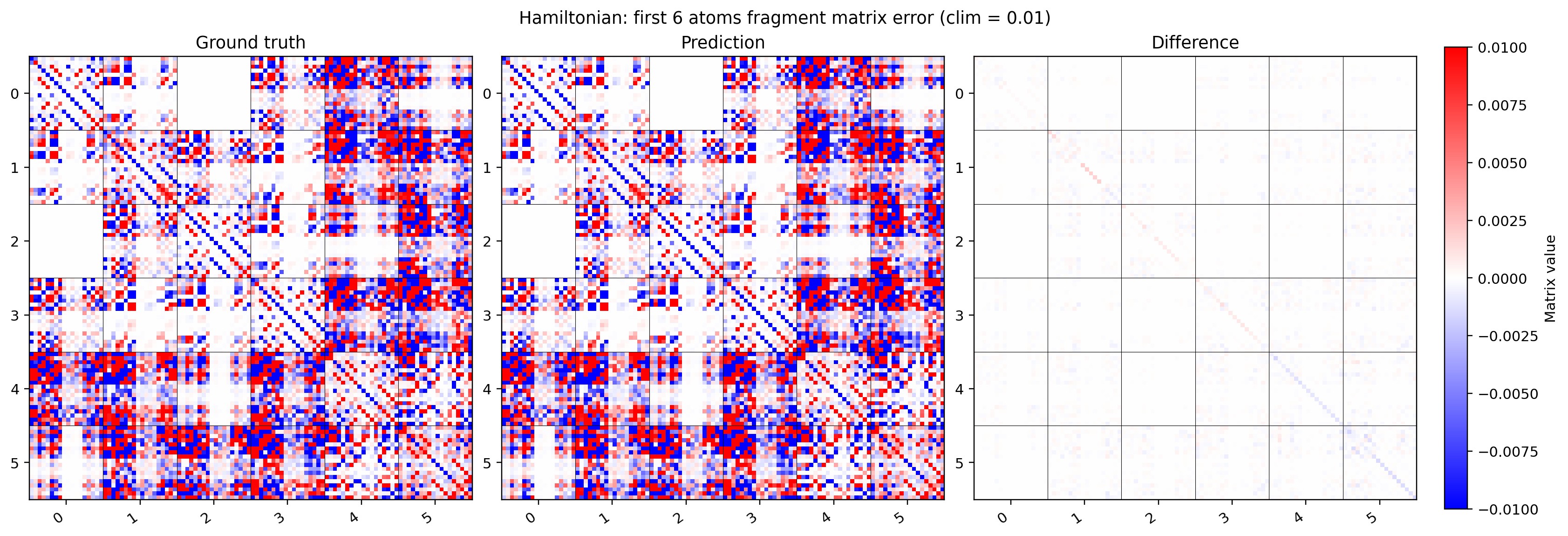}
\caption{Hamiltonian fragments for the first six atoms of a representative ZnCu$_2$Sn(SeS)$_2$ validation structure. The common color range is $\pm0.01$~eV.}
\label{fig:zncusnses-hamiltonian}
\end{figure}

\begin{figure}[H]
\centering
\includegraphics[width=0.82\textwidth]{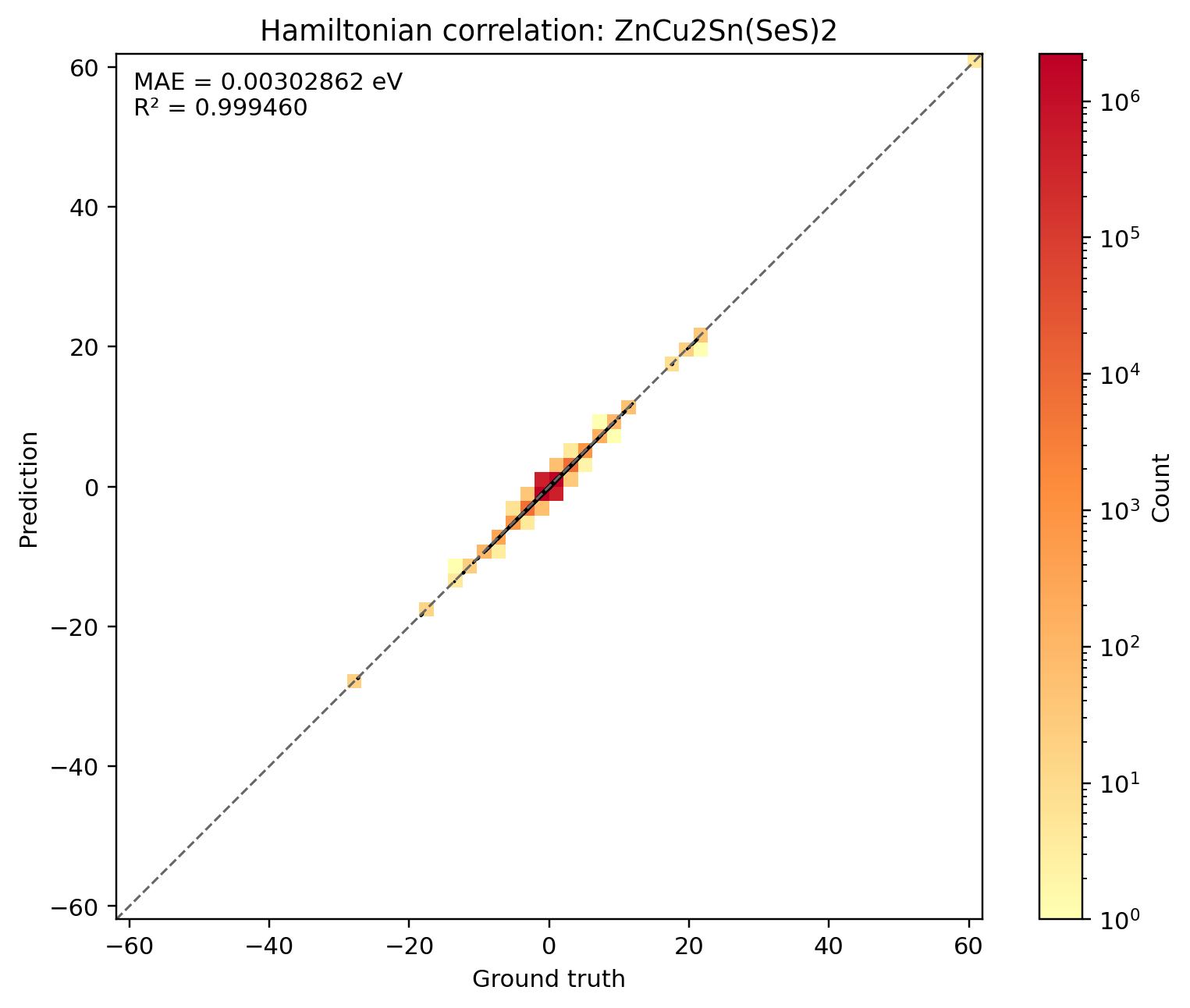}
\caption{Hamiltonian-element correlation for the representative ZnCu$_2$Sn(SeS)$_2$ validation structure. The density heatmap uses a logarithmic count scale, with individual matrix elements overlaid as black points.}
\label{fig:zncusnses-correlation}
\end{figure}

Figure~\ref{fig:zncusnses-hamiltonian} resolves a representative prediction into atom-pair matrix blocks. The $\pm0.01$~eV range is narrow compared with the Hamiltonian elements and exposes residual structure in both on-site and off-site blocks.
The Hamiltonian correlation in Fig.~\ref{fig:zncusnses-correlation} includes all 5\,345\,640 shift-resolved matrix elements of the same validation structure. Its structure-level MAE is 0.003029~eV and $R^2=0.999460$, consistent with the aggregate validation error despite the chemically and orbitally heterogeneous block structure.

\subsection{Electronic density of states of SiO$_2$}

The SiO$_2$ dataset consists of 150-atom glass structures. Variations in bond lengths, bond angles, and coordination environments replace the small set of symmetry-equivalent neighborhoods present in a crystal.
This makes the dataset a test of whether the local equivariant representation remains accurate across a broad distribution of amorphous environments.
Table~\ref{tab:sio2-validation} reports the aggregate validation metrics. For this Hamiltonian-only model, the band energy is evaluated by contracting the predicted Hamiltonian with the reference density.
Its compact configuration is provided in \ref{app:model-config-sio2}.

\begin{table}[H]
\centering
\caption{Validation metrics for the glassy SiO$_2$ model.}
\label{tab:sio2-validation}
\begin{tabular}{lc}
\toprule
Metric & Value (relative) \\
\midrule
Hamiltonian MAE & 0.001996~eV (1.67\%) \\
Band-energy MAE & 0.146~eV/atom (0.193\%) \\
\bottomrule
\end{tabular}
\end{table}

\begin{figure}[H]
\centering
\includegraphics[width=0.88\textwidth]{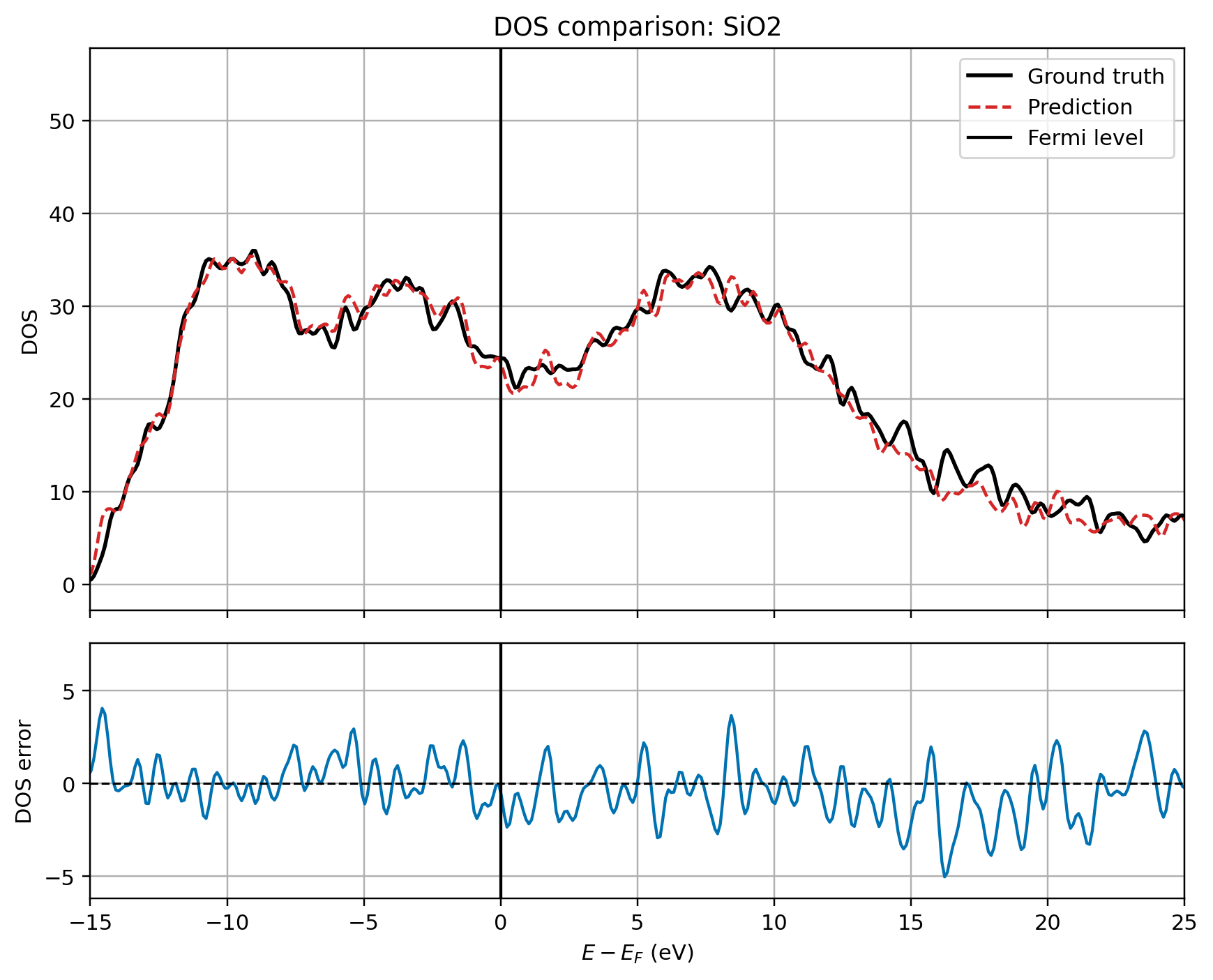}
\caption{Density of states and its prediction error for a representative glassy SiO$_2$ validation structure. The calculation uses Gaussian broadening with $\sigma=0.2$~eV on a $2\times2\times2$ k-point mesh. The solid black line marks the Fermi level.}
\label{fig:sio2-dos}
\end{figure}

The density of states in Fig.~\ref{fig:sio2-dos} tests the collective spectral consequence of the predicted matrix rather than individual matrix elements.

\subsection{Band structure, density of states, and density matrix of silicon}

The silicon dataset retains the diamond-crystal topology while introducing small atomic displacements ($\sigma = 0.02$~\r{A}). Its single-species composition and regular coordination remove the chemical and configurational complexity of the other two systems, allowing the attainable precision of the equivariant model to be assessed on a simpler crystalline electronic structure.
The aggregate validation metrics of the joint Hamiltonian--density--overlap model are given in Table~\ref{tab:silicon-validation}; its compact configuration is provided in \ref{app:model-config-silicon}.

\begin{table}[H]
\centering
\caption{Validation metrics for the selected silicon model.}
\label{tab:silicon-validation}
\begin{tabular}{lc}
\toprule
Metric & Value (relative) \\
\midrule
Hamiltonian MAE & $4.352\times10^{-4}$~eV (0.355\%) \\
Density-matrix MAE & $5.758\times10^{-5}$ (1.95\%) \\
Overlap-matrix MAE & $1.390\times10^{-5}$ (0.109\%) \\
Band-energy MAE & 0.0207~eV/atom (0.0631\%) \\
\bottomrule
\end{tabular}
\end{table}

\begin{figure}[H]
\centering
\includegraphics[width=\textwidth]{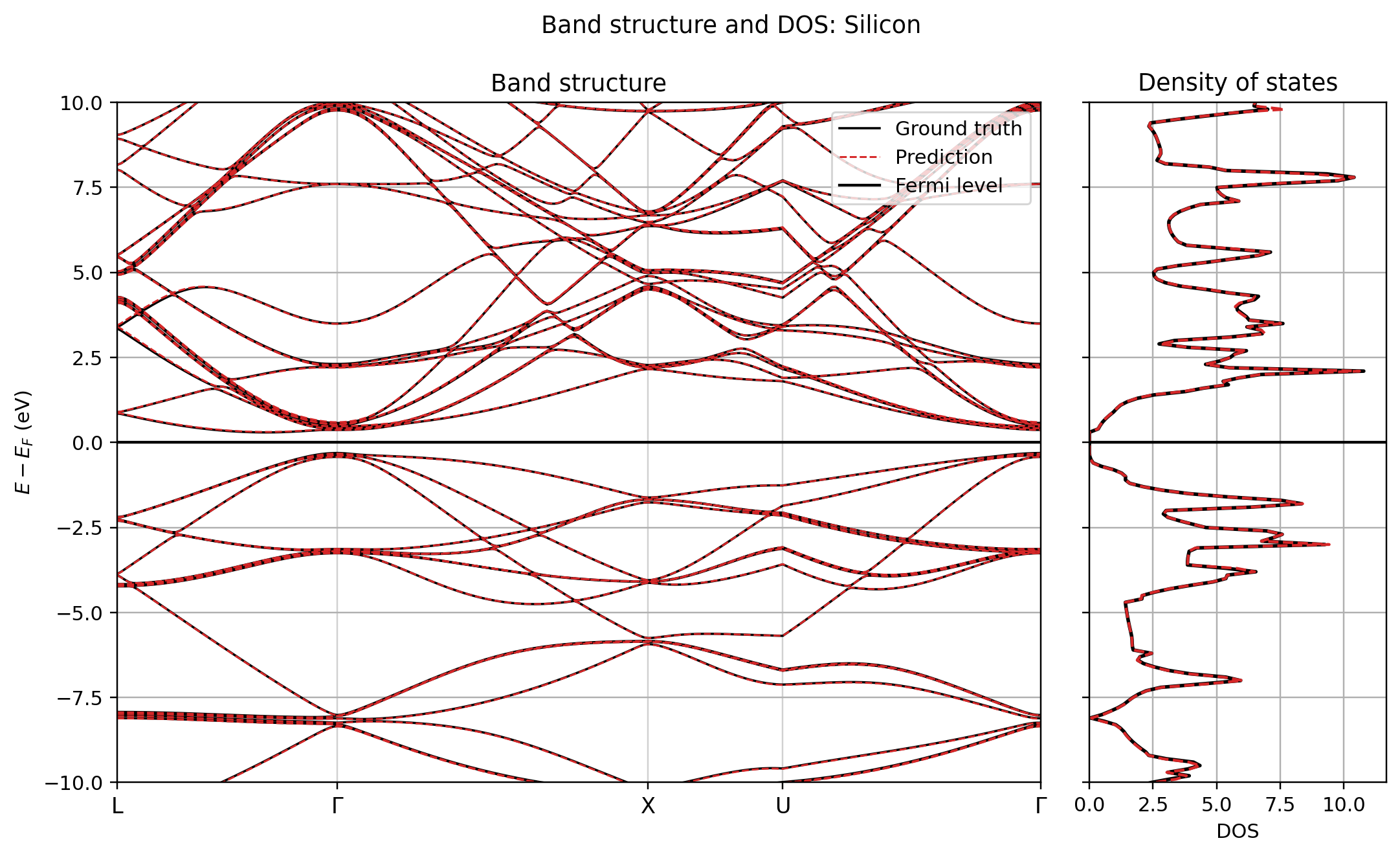}
\caption{Band structure and density of states for a representative silicon validation structure. The path is $L\!\rightarrow\!\Gamma\!\rightarrow\!X\!\rightarrow\!U\!\rightarrow\!\Gamma$, and the DOS uses tetrahedron integration on a $4\times4\times4$ k-point mesh. The solid black line marks the Fermi level. The model is able to almost perfectly reproduce the spectral features of the reference calculation on this simpler dataset.}
\label{fig:silicon-bands-dos}
\end{figure}

The predicted Hamiltonian combined with the reference overlap determines the bands and DOS in Fig.~\ref{fig:silicon-bands-dos}. For the representative structure, the resulting operators reproduce the displayed occupied and low-lying unoccupied bands.
The valence-band maximum and conduction-band dispersion around the high-symmetry points are reproduced without fitting the bands as direct targets. Tests of the physical constraints of the predicted operators, including the positive definiteness of the predicted overlap, are reported in \mbox{\ref{app:physical-constraints}}.

\begin{figure}[H]
\centering
\includegraphics[width=\textwidth]{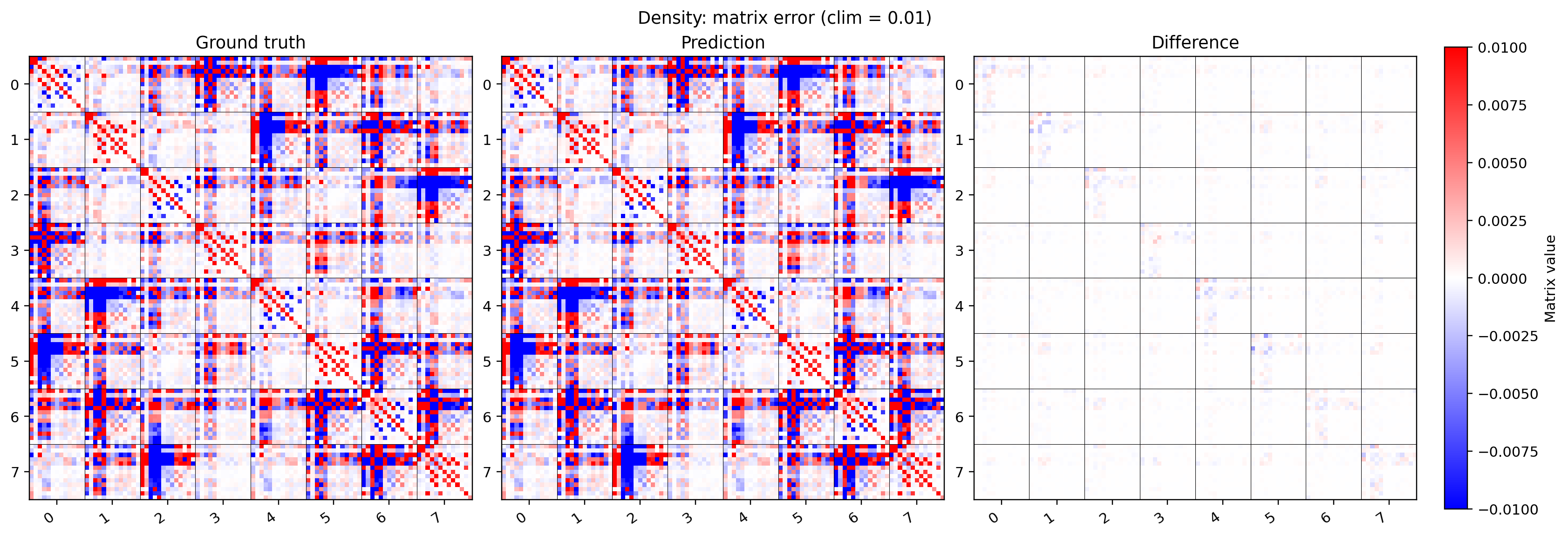}
\caption{The central density matrix of the eight-atom representative silicon validation structure. Each atom-pair contribution is an $n\times n$ local-orbital block, with $n=13$ for this calculation. The common color range is $\pm0.01$.}
\label{fig:silicon-density}
\end{figure}

\begin{figure}[H]
\centering
\begin{minipage}[t]{0.49\textwidth}
\centering
\includegraphics[width=\linewidth]{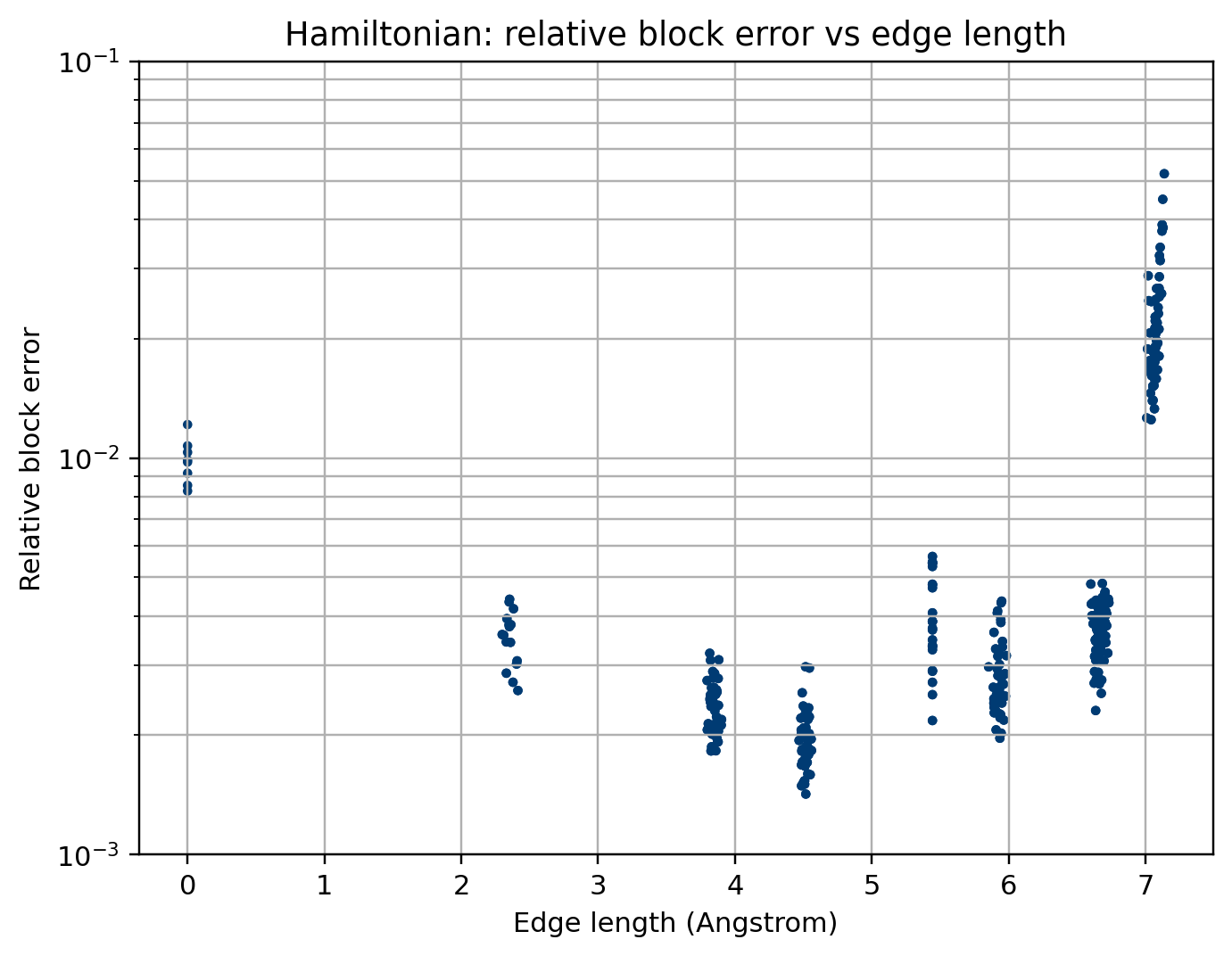}
\small (a) Hamiltonian
\end{minipage}\hfill
\begin{minipage}[t]{0.49\textwidth}
\centering
\includegraphics[width=\linewidth]{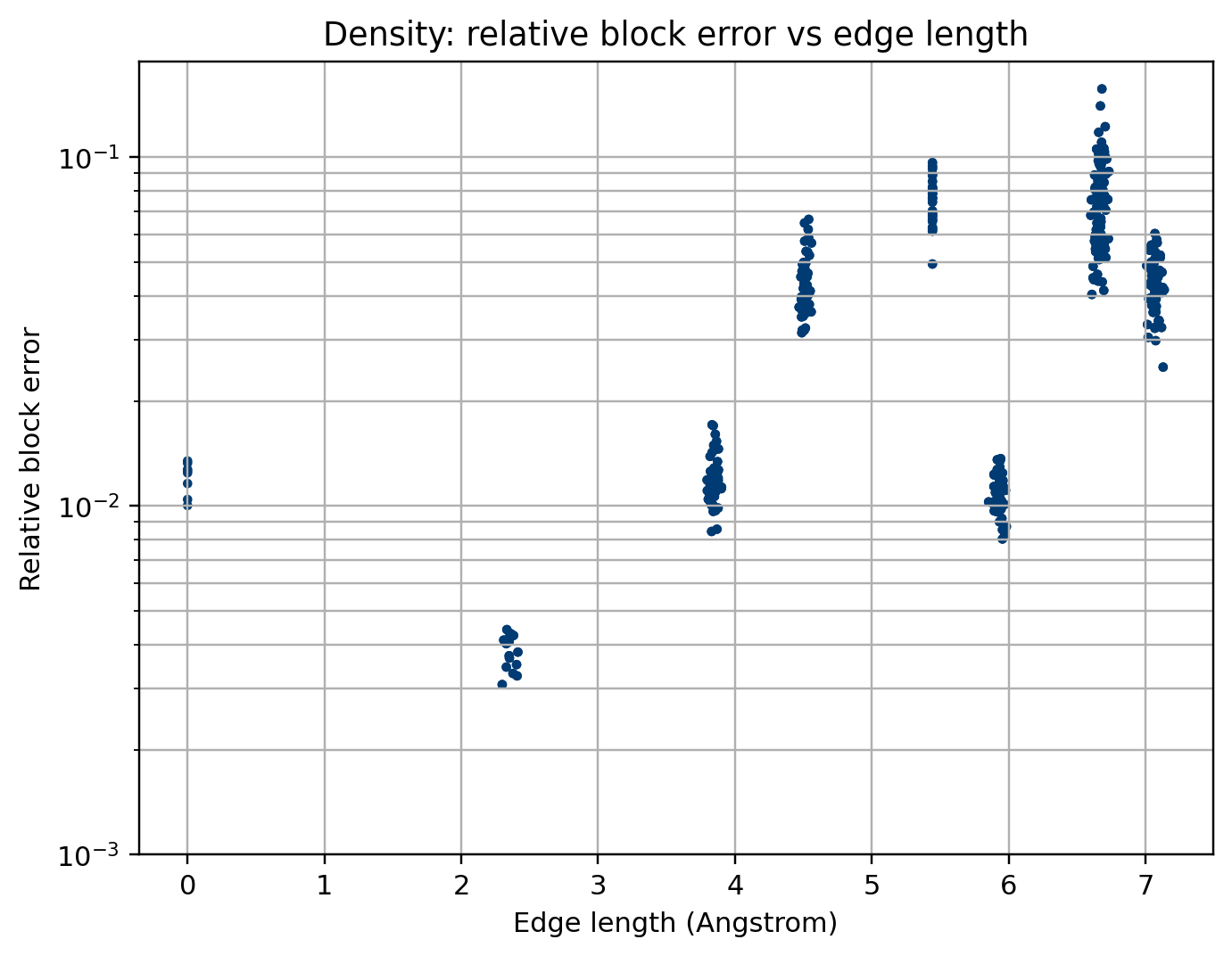}
\small (b) Density
\end{minipage}
\caption{Relative block error by atom-pair distance for the representative silicon structure. Each point is one shift-resolved block.}
\label{fig:silicon-density-errors}
\end{figure}

For each atom pair and periodic shift, a Hamiltonian or density contribution is an $n\times n$ block in the local orbital basis, where $n$ is the number of orbitals associated with each atom for the chosen basis. Figure~\ref{fig:silicon-density} compares the central-cell reference and predicted density matrices of the representative structure and shows their element-wise error.
For a predicted block $M^{\mathrm{pred}}$ and its reference $M^{\mathrm{ref}}$, the relative block error shown in Fig.~\ref{fig:silicon-density-errors} is
\begin{equation}
\epsilon_{\mathrm{rel}}(M^{\mathrm{pred}},M^{\mathrm{ref}})
=\frac{\frac{1}{n^2}\sum_{a,b}|M^{\mathrm{pred}}_{ab}-M^{\mathrm{ref}}_{ab}|}
{\max\!\left(\frac{1}{n^2}\sum_{a,b}|M^{\mathrm{ref}}_{ab}|,10^{-12}\right)}.
\end{equation}
The discrete distance groups correspond to coordination shells of the perturbed crystal.
Hamiltonian blocks have their smallest relative errors over the strongly coupled intermediate shells and larger relative errors near the cutoff, where normalization by weak target blocks magnifies the ratio.
The density matrix shows greater shell-to-shell variation because its blocks do not fall off with magnitude as quickly as the Hamiltonian blocks do.

\section{Ablations}
\label{sec:ablations}

\texttt{Mandala}'s broad configuration system exposes data transformations, model
components, and training objectives through a common interface, which makes
controlled ablations straightforward to execute. The studies below isolate
several of these choices while keeping the remaining configuration fixed.
Their results demonstrate benefits for the specific datasets, architectures,
and training regimes considered here; they do not establish that any one
setting is universally superior.

\subsection{Energy guidance}

\begin{figure}[H]
\centering
\includegraphics[width=\textwidth]{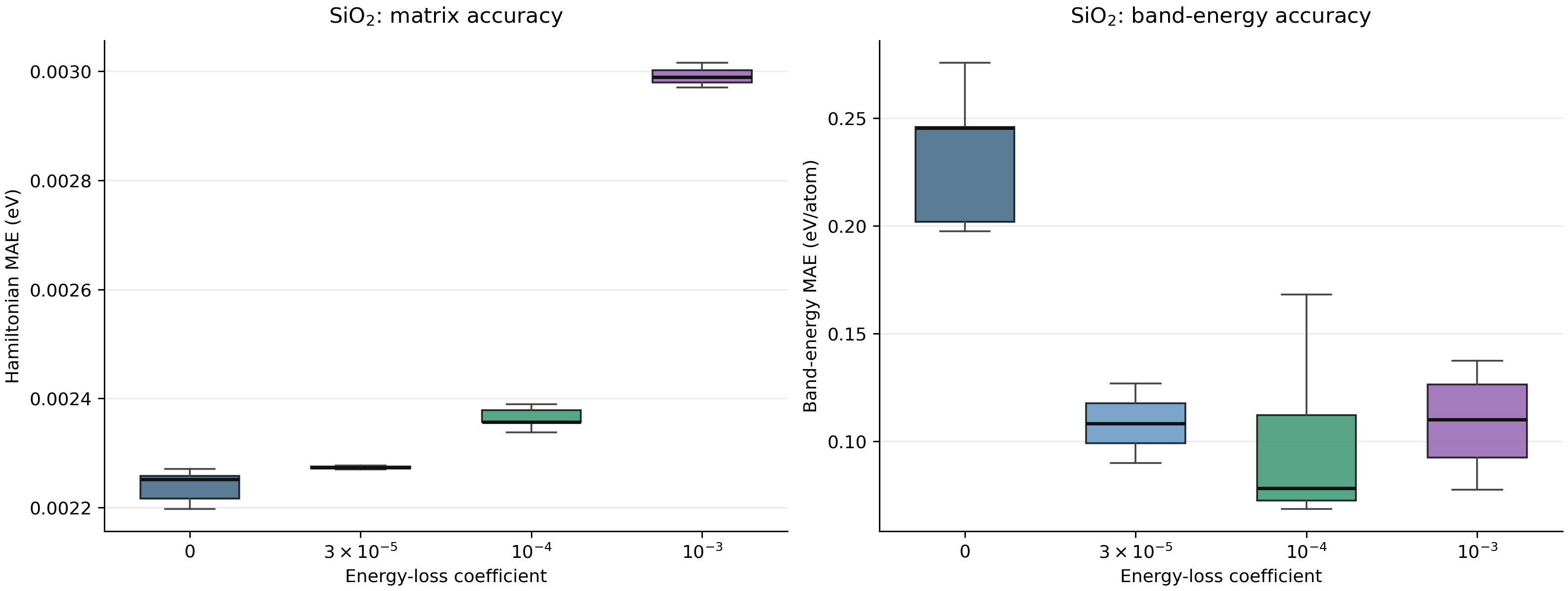}
\caption{Held-out test Hamiltonian and per-atom band-energy MAE for five matched SiO$_2$ seeds with energy guidance.}
\label{fig:result-siox-energy-guidance}
\end{figure}

The presented held-out test results show that energy guidance introduces an explicit tradeoff between element-wise
Hamiltonian accuracy and the aggregate band-energy observable. As shown in
Fig.~\ref{fig:result-siox-energy-guidance}, a loss coefficient of $10^{-4}$
reduces the median band-energy error from $0.245$ to $0.078$~eV/atom, a
$68\%$ improvement, while increasing the median Hamiltonian MAE by only
$4.7\%$. Increasing the coefficient to
$10^{-3}$ is detrimental to both metrics relative to $10^{-4}$, indicating
that excessive observable weighting diverts optimization away from an
effective joint solution.

\subsection{Spectral guidance}

Spectral guidance acts directly on eigenvalues in the energy window of
interest. Across ten matched ZnCu$_2$Sn(SeS)$_2$ seeds, it reduces the median
held-out test spectral MAE from $0.579$ to $0.0391$~eV. The paired mean
relative improvement is $92.1\%$, and all ten seeds improve. The Hamiltonian
effect is near neutral: the median remains $0.002791$~eV and the paired mean
improvement is $0.51\%$, with a 95\% confidence interval spanning improvement
and degradation.

\begin{figure}[H]
\centering
\includegraphics[width=\textwidth]{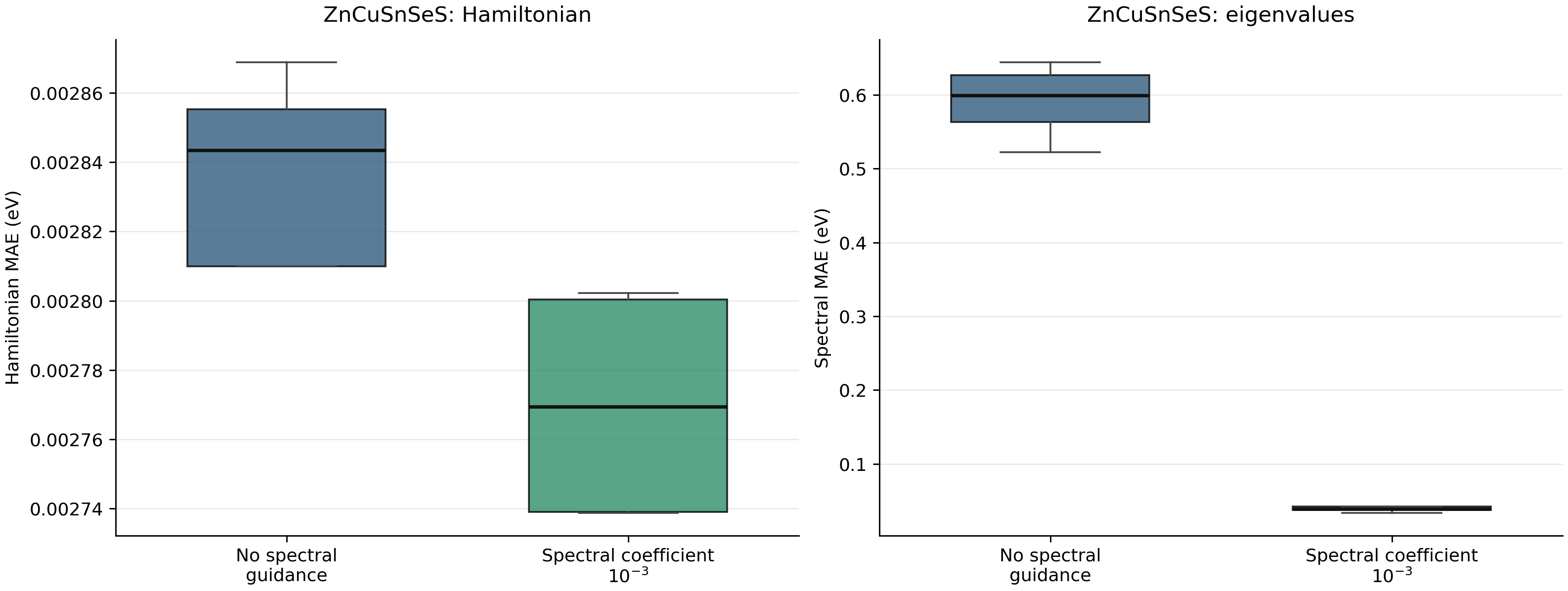}
\caption{Held-out test Hamiltonian and eigenvalue MAE for ten matched ZnCu$_2$Sn(SeS)$_2$ seeds with spectral guidance.}
\label{fig:result-zncusnses-spectral-guidance}
\end{figure}

\subsection{Distance-envelope factorization and node-message aggregation}

The fitted distance envelope supplies an explicit pair-dependent radial prior
to the Hamiltonian head. Prediction factorization lowers the median test
Hamiltonian MAE by $3.8\%$, with improvements in four of five matched seeds.
The effect is modest but reasonably consistent across this ablation.

The choice of node-message aggregation has a larger influence. Replacing
average aggregation with attention lowers the median silicon
validation Hamiltonian MAE by $10.1\%$.
The evaluation was performed on the validation split instead of a held-out test set in this case.

\begin{figure}[H]
\centering
\begin{minipage}[t]{0.49\textwidth}
\centering
\includegraphics[width=\linewidth]{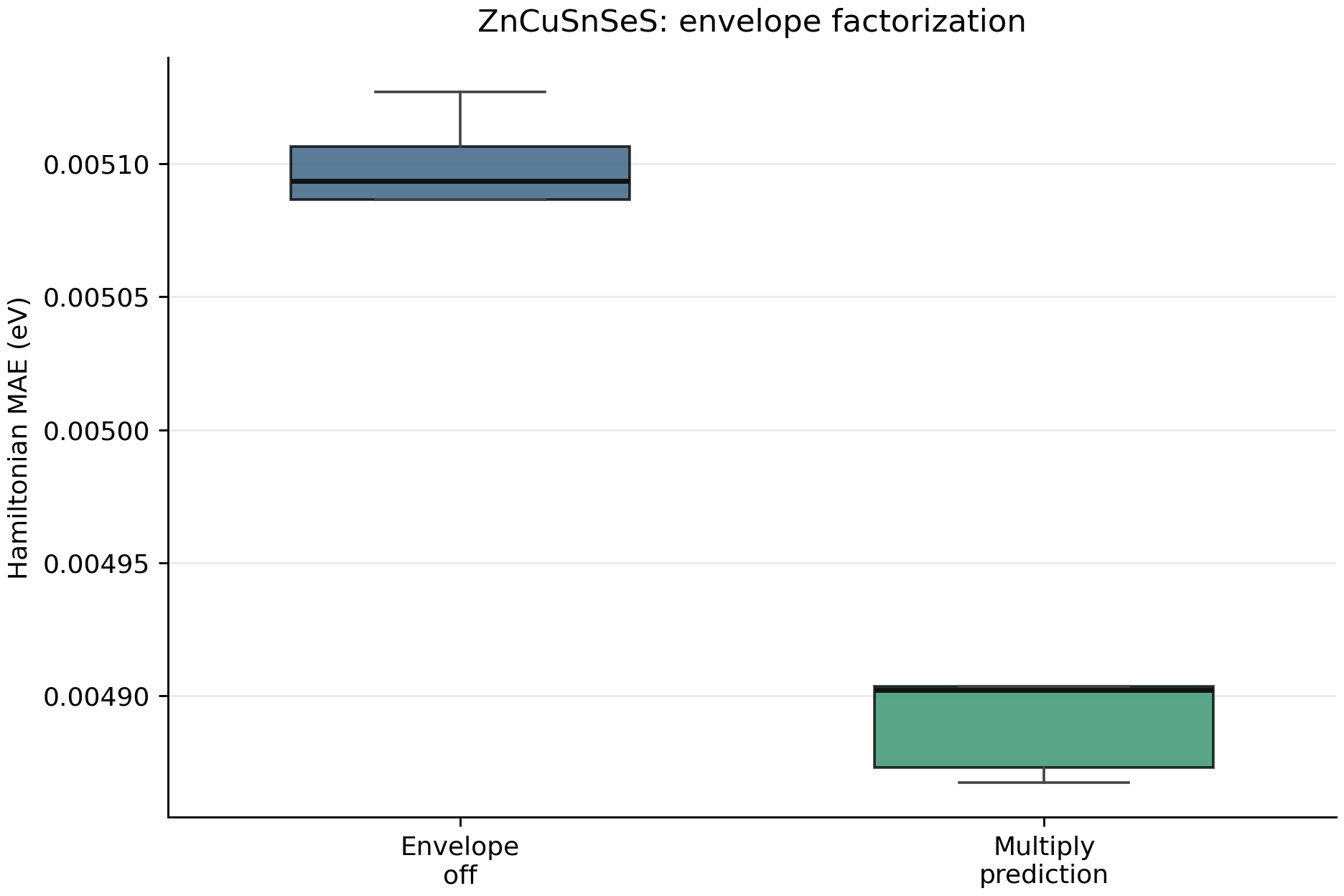}
\small (a) Distance-envelope factorization
\end{minipage}\hfill
\begin{minipage}[t]{0.49\textwidth}
\centering
\includegraphics[width=\linewidth]{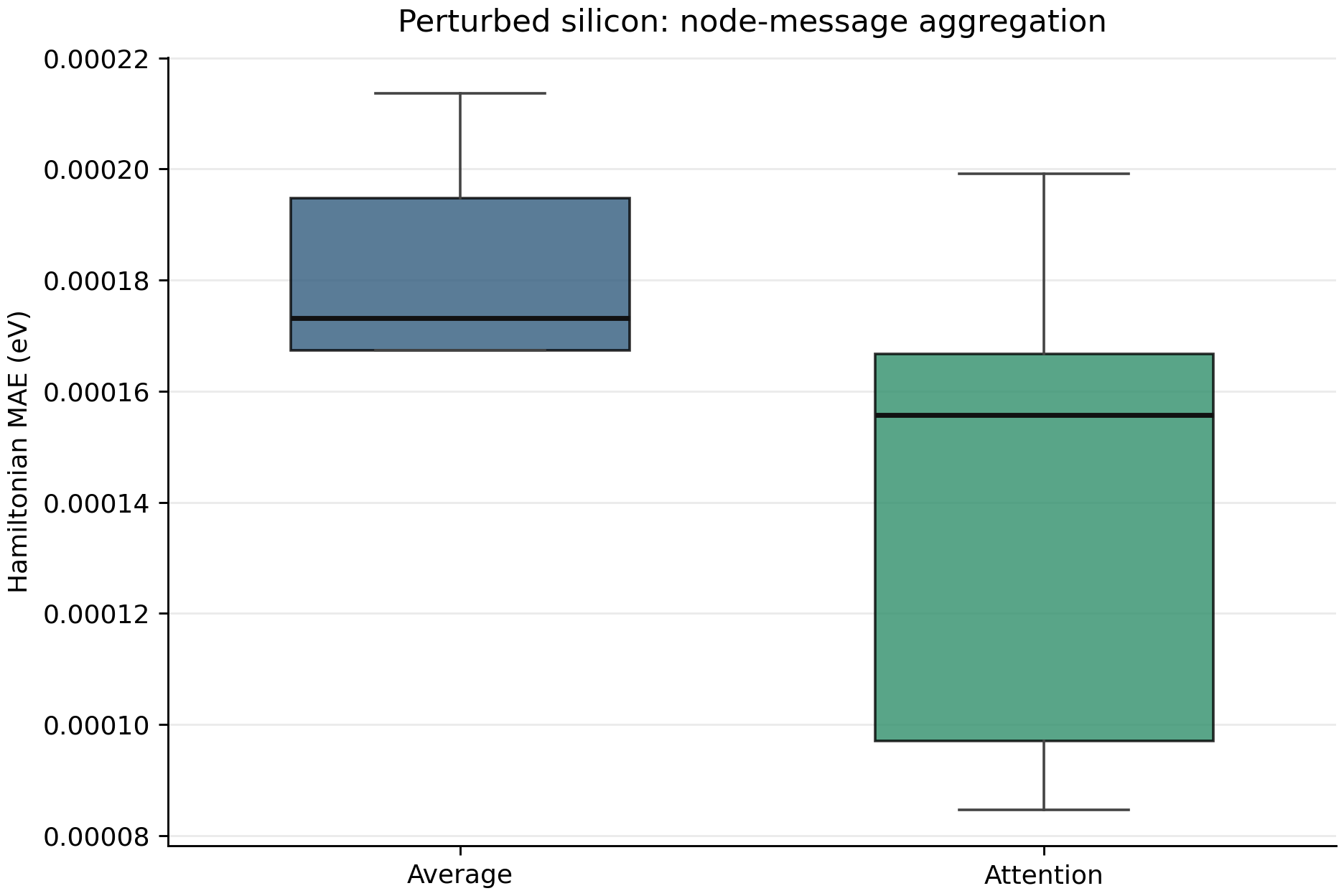}
\small (b) Node-message aggregation
\end{minipage}
\caption{Hamiltonian MAE for distance-envelope factorization on ZnCu$_2$Sn(SeS)$_2$ and node-message aggregation on silicon.}
\label{fig:result-matrix-ablations}
\end{figure}

\section{Discussion}
\label{sec:discussion}

The three demonstrations establish that one sparse-operator pipeline can accommodate materially different regimes: a five-element periodic compound with many chemical-pair channels, a 150-atom amorphous oxide with broad local structural variation, and a crystalline system in which Hamiltonian, density, and overlap matrices are learned jointly. The downstream figures also show that the learned real-space operators can be carried into matrix-resolved diagnostics, density-of-states calculations, and band-structure calculations without introducing a separate observable-prediction model.

These results do not establish state-of-the-art accuracy. The showcased models are selected validation checkpoints and the representative figures visualize individual validation structures. Their purpose is to verify the breadth and composability of the software, not a definitive benchmark. The seed-matched ablations answer narrower questions about configured components and loss terms. In particular, the spectral study demonstrates that an operator-derived objective can substantially improve eigenvalues in a chosen energy window while leaving the matrix metric nearly unchanged; it does not imply that spectral guidance has the same effect for every dataset or training regime.

The software-level contribution is the shared route from backend parsing through sparse periodic block alignment, symmetry conversion, configurable equivariant learning, differentiable sparse observables, and electronic-structure reporting. Keeping these stages behind common data and model interfaces makes it possible to change targets or architecture components without constructing a new end-to-end code path. The current boundaries remain important: the demonstrations use real non-spin-polarized operators in fixed localized bases, dense diagonalization remains a downstream scaling limit, and the learned operators do not yet satisfy every many-body or density-matrix consistency condition by construction.

\clearpage
\section{Public Interface and Usage Examples}
\label{sec:usage-examples}

The following minimal examples illustrate snapshot loading, model restoration, sparse contractions, and band-structure postprocessing. They are run from the repository root with \texttt{src} on the Python path; runnable versions are provided in \texttt{examples/quickstart}. A calculation can be loaded and its sparse operator-derived quantities examined directly:

\begin{lstlisting}
from data.snapshot import Snapshot

snapshot = Snapshot.from_openmx(
    "sample.matrix", "sample.info.out", convention="e3nn"
)
print(snapshot.get_energy())
print(snapshot.get_number_of_electrons())
\end{lstlisting}

A trained model is restored with its saved configuration and orbital metadata, then applied directly to the same snapshot:

\begin{lstlisting}
import torch
from net.evaluation import predict_snapshot, restore_model_for_snapshot

device = "cuda" if torch.cuda.is_available() else "cpu"
model = restore_model_for_snapshot("best_model.pt", snapshot, device=device)
predictions = predict_snapshot(model, snapshot)
predicted_hamiltonian = predictions["hamiltonian"]
\end{lstlisting}

One can evaluate a Hamiltonian-only model on band energy using the reference density:

\begin{lstlisting}
from core.sparse_math import trace_matmul_sparse_block_matrix

band_energy = trace_matmul_sparse_block_matrix(
    predicted_hamiltonian, snapshot.density
)
\end{lstlisting}

Likewise, the predicted Hamiltonian can be combined explicitly with a reference overlap matrix for generalized-eigenvalue band-structure analysis:

\begin{lstlisting}
from net.evaluation import predictions_to_snapshot

predicted = predictions_to_snapshot(
    predictions, snapshot, reference_matrices=("overlap", "density")
)
bands = predicted.get_band_structure(
    path="GXMGR",
    npoints=200,
)
print(bands.eigenvalues.shape)
\end{lstlisting}

This interface supports a wide range of workflows, with minimal user code required to load data, restore models, and evaluate observables. New layers and training objectives are added in the same way, without modifying the library; worked examples are described in \mbox{\ref{app:extending}}.

\section{Conclusion}
\label{sec:conclusion}

\texttt{Mandala} is a scientific software framework for learning sparse electronic-structure operators with E(3)-equivariant graph neural networks. Its defining features are the explicit sparse block representation of Hamiltonian, overlap, and density matrices; the modular software layers that connect parsing, basis harmonization, irrep mapping, graph construction, model definition, and training; and the ability to supervise not only matrices themselves but also observables derived from them.

The framework combines capabilities that are often developed separately. From the electronic-structure side, it retains access to operator-level information and basis-aware quantities. From the atomistic machine-learning side, it inherits the flexibility of modern equivariant message passing, multitask learning, and automated hyperparameter search. By combining these perspectives, \texttt{Mandala} supports workflows in which one can train models on matrix, band-energy, electron-count and spectral fidelity within one coherent implementation.

In this sense, \texttt{Mandala} is aimed at simulation settings that go beyond conventional machine-learning interatomic potentials by retaining access to the predicted electronic structure, including Hamiltonian-derived quantities such as band structure and related observables. This makes the framework particularly relevant for problems in which explicitly electronic effects, such as charge transfer or electronically driven changes in bonding, are part of the physics that must be modeled rather than treated only indirectly. In its current form, \texttt{Mandala} is positioned as a modular platform for operator-level machine learning in electronic-structure simulation and for systematic exploration of equivariant architectures built on that operator-centered representation.

The capabilities of \texttt{Mandala} for predicting the Hamiltonian matrix, density matrix, density of states, and band structure are illustrated for a variety of systems including the chalcogenide ZnCu$_2$Sn(SeS)$_2$, amorphous silicon oxide, and crystalline silicon.

\section*{Acknowledgments}

This work was supported by the Center for Advanced Systems Understanding (CASUS), which is financed by Germany’s Federal Ministry of Research, Technology and Space (BMFTR) and by the Saxon State government out of the State budget approved by the Saxon State Parliament, and by the University of Wroc\l aw. Computational resources were provided by the \texttt{hemera} and \texttt{rosi} clusters of HZDR.

\section*{Data and Model Availability}

The datasets used in this study and the trained models evaluated in this publication are available through the RODARE repository at \url{https://doi.org/10.14278/rodare.4816}. The workflows to train and evaluate these models are described in \mbox{\ref{app:reproduction}}.

\paragraph*{\bf Statement on the Use of Generative AI}
\vspace{1cm}
During the preparation of this work the authors used generative AI in order to check the text for grammar mistakes and make improvements to the overall language. After using this service, the authors reviewed and edited the content as needed and take full responsibility for the content of the publication.

\clearpage
\bibliographystyle{elsarticle-num}
\bibliography{references}

\clearpage
\appendix
\section{Compact configurations of the evaluated models}
\label{app:model-configurations}

The three selected models share two E(3)-equivariant message-passing layers, separate-weight tensor products, a rich edge encoder with spherical-harmonic tensor-square features, invariant attention for node aggregation, species-pair-split output heads, separate shifted-self handling, MAE matrix training, and no spectral loss. Table~\ref{tab:evaluated-configs} lists the settings that distinguish the models. Complete machine-readable resolved configurations accompany the manuscript in \texttt{../data/evaluated\_model\_configs/}; paths and experiment-tracking metadata are omitted from the paper.

\begin{table}[H]
\centering
\scriptsize
\caption{Distinguishing settings of the three models used in the capability demonstrations. H, D, and S denote Hamiltonian, density, and overlap targets.}
\label{tab:evaluated-configs}
\begin{tabular}{p{0.18\textwidth}p{0.22\textwidth}p{0.22\textwidth}p{0.22\textwidth}}
\toprule
Setting & ZnCu$_2$Sn(SeS)$_2$ & SiO$_2$ & Perturbed Si \\
\midrule
Appendix label &
\ref{app:model-config-zncusnses} &
\ref{app:model-config-sio2} &
\ref{app:model-config-silicon} \\
Targets & H & H & H, D, S \\
Cutoff / $\ell_{\max}$ & 11~\r{A} / 6 & 10~\r{A} / 4 & 8~\r{A} / 4 \\
Neck / head depth & 2 / 1 & 1 / 1 & 2 / 2 \\
Head nonlinearity & NormAct & NormAct & FiLM \\
Radial envelope & Multiply prediction & Multiply prediction & Off \\
Energy guidance & Off & Off & $3\times10^{-3}$ \\
Learning rate & $5\times10^{-3}$ & $2\times10^{-4}$ & $3\times10^{-4}$ \\
Training seed & 44 & 42 & 43 \\
Maximum wall time & 47.0~h & 47.25~h & 47.25~h \\
\bottomrule
\end{tabular}
\end{table}

\subsection{ZnCu$_2$Sn(SeS)$_2$ Hamiltonian model}
\label{app:model-config-zncusnses}

The model uses the hidden representation
\begin{align*}
&384{\times}0e+48{\times}0o+24{\times}1e+192{\times}1o\\
&\quad+72{\times}2e+24{\times}2o+24{\times}3e+72{\times}3o\\
&\quad+48{\times}4e+12{\times}4o+12{\times}5e+36{\times}5o+24{\times}6e.
\end{align*}
Its fitted pair-dependent radial envelope multiplies the predicted Hamiltonian blocks. The full resolved configuration is available in supplementary materials.

\subsection{SiO$_2$ Hamiltonian model}
\label{app:model-config-sio2}

The model uses the hidden representation
\[
128{\times}0e+64{\times}0o+32{\times}1e+32{\times}1o+
24{\times}2e+24{\times}2o+16{\times}3e+16{\times}3o+8{\times}4e.
\]
Its fitted pair-dependent radial envelope multiplies the predicted Hamiltonian blocks. The full resolved configuration is available in supplementary materials.

\subsection{Silicon joint operator model}
\label{app:model-config-silicon}

The model uses the hidden representation
\[
128{\times}0e+128{\times}0o+64{\times}1e+64{\times}1o+
32{\times}2e+32{\times}2o+16{\times}3e+16{\times}3o+16{\times}4e.
\]
It jointly predicts Hamiltonian, density, and overlap matrices without a radial output envelope. Band-energy guidance contracts the predicted Hamiltonian with the reference density during training. The full resolved configuration is available in supplementary materials.

\section{Error distributions of the evaluated models}
\label{app:error-distributions}

Table~\ref{tab:error-distributions} complements the mean absolute errors in Tables~\ref{tab:zncusnses-validation}--\ref{tab:silicon-validation} with the root-mean-square error (RMSE), the 95th and 99th percentiles ($p_{95}$, $p_{99}$) and the maximum of the absolute matrix-element error, and the coefficient of determination $R^2$. All statistics are pooled over every matrix element of the validation structures of each model.
For the three Hamiltonian models, 99\% of the matrix elements have absolute errors below 33.5, 26.0, and 5.40~meV for ZnCu$_2$Sn(SeS)$_2$, SiO$_2$, and silicon, respectively, which is 11--13 times the respective MAE, and $R^2$ exceeds 0.999 in every case. The maximum errors are 8--40 times larger than $p_{99}$, reaching 1.1~eV for a single SiO$_2$ matrix element, but they are confined to the top 1\% of the error distribution. The density and overlap matrices of the silicon model show the same pattern.

\begin{table}[H]
\centering
\caption{Distribution of the absolute matrix-element errors on the validation structures (number of structures in parentheses). Hamiltonian errors are in meV; density and overlap errors are dimensionless.}
\label{tab:error-distributions}
\resizebox{\linewidth}{!}{%
\begin{tabular}{llrrrrrr}
\toprule
Model & Operator & MAE & RMSE & $p_{95}$ & $p_{99}$ & max & $R^2$ \\
\midrule
ZnCu$_2$Sn(SeS)$_2$ (20) & $H$ & 3.012 & 7.49 & 13.7 & 33.5 & 278 & 0.9995 \\
SiO$_2$ (10) & $H$ & 1.996 & 6.79 & 10.5 & 26.0 & 1106 & 0.9999 \\
Silicon (16) & $H$ & 0.435 & 1.61 & 1.96 & 5.40 & 78.5 & 0.99998 \\
Silicon (16) & $D$ & $5.76 \times 10^{-5}$ & $1.41 \times 10^{-4}$ &
$2.14 \times 10^{-4}$ & $5.58 \times 10^{-4}$ &
$1.19 \times 10^{-2}$ & 0.99991 \\
Silicon (16) & $S$ & $1.39 \times 10^{-5}$ & $4.76 \times 10^{-5}$ &
$7.93 \times 10^{-5}$ & $1.86 \times 10^{-4}$ &
$3.87 \times 10^{-3}$ & 0.999999 \\
\bottomrule
\end{tabular}%
}
\end{table}

\section{Extending MANDALA}
\label{app:extending}

New network components are added without modifying the library. A layer consists of a PyTorch module and a factory function that is registered under a name with \texttt{register\_transform}. The factory checks the contract of the inputs (domain, invariance, irreps) and declares the irreps of the outputs. Once the defining module is imported, the layer can be used by name in any YAML recipe:

\begin{lstlisting}
@register_transform("scale_by_length")
def build(options, inputs, mapper):
    values, lengths = inputs
    if values.domain != "edge" or values.irreps is None:
        raise ValueError("expects irrep-valued edge features")
    return ScaleByLength(**options), [values]
\end{lstlisting}

\begin{lstlisting}[language={}]
- id: damp_edges
  transform: scale_by_length
  inputs: [edge_1, edge_lengths]
  outputs: [edge_1_damped]
  options: {hidden: 16}
\end{lstlisting}

The documentation\footnote{\url{https://mandala-org.github.io/Mandala/extending/index.html}} uses this mechanism to implement a new layer for each of the twelve layer kinds of the network: radial basis, node and edge encoders, convolution, aggregation, nonlinearity, normalization, MLP block, residual update, edge update, readout, and output envelope. Examples include a Bessel radial basis, an attention aggregation with a learned softmax temperature, and a learned radial envelope for each atom pair. The twelve layers are assembled into a complete network that is defined only in YAML, shares no layers with the default model apart from the spherical harmonics and the operator decoder, and is trained on the bundled silicon data. A verification script checks numerically that each new layer and the assembled model are E(3)-equivariant; the largest deviation of the predicted Hamiltonian under rotations, improper rotations, and inversion is $3\times10^{-10}$.

\begin{sloppypar}
New observables are added in the same way. A differentiable function of the predicted and reference operators is registered with \texttt{register\_objective\_term} and enabled with a weight in the YAML configuration, without changes to the network or the training loop. The documentation demonstrates this with the mean onsite Hamiltonian trace per atom.
\end{sloppypar}

\section{Reference architectures and inference efficiency}
\label{app:reference-architectures}

\begin{sloppypar}
DeepH, DeepH-E3, and MACE-H are provided as reference configurations (\texttt{configs/reference/\{deeph,deeph\_e3,mace\_h\}.yaml}) assembled from MANDALA's layer library. Their implementations reproduce the original codes to numerical precision. For pinned versions of the original codes, the test suite checks that the predictions, the gradients, and the parameters after each of three Adam optimizer steps agree to a relative tolerance of $10^{-8}$ in double precision.
\end{sloppypar}

Table~\ref{tab:inference-efficiency} compares the inference time of these implementations with the original codes on graphene and MoS$_2$. Times are medians over repeated evaluations of three structures per system, with randomly initialized weights and one structure per batch, measured with PyTorch~2.7.1 and e3nn~0.6.0 on one NVIDIA H100 GPU. For a like-for-like comparison of the networks, the native codes use the same vectorized host-side data handling as MANDALA; as distributed, they take 406--2011\,ms per structure. In eager mode, MANDALA is within 4--7\% of the native implementations. Compilation with \texttt{torch.compile} and CUDA graphs reduces the inference time of DeepH-E3 and MACE-H by a factor of 1.7--2.9. Compilation takes about 4--6 minutes and is specific to a maximum number of atoms and edges, so it pays off for large numbers of similarly sized structures, for example along a molecular dynamics trajectory.

\begin{table}[H]
\centering
\caption{Median end-to-end inference time per structure (ms) on one NVIDIA H100 GPU. Percentages are relative to the native implementation.}
\label{tab:inference-efficiency}
\resizebox{\linewidth}{!}{%
\begin{tabular}{llrrr}
\toprule
Architecture & System & Native & MANDALA (eager) & MANDALA (compiled) \\
\midrule
DeepH    & Graphene & 95  & 101 (+7\%) & 109 (+15\%) \\
         & MoS$_2$  & 307 & 318 (+4\%) & 323 (+5\%) \\
DeepH-E3 & Graphene & 269 & 279 (+4\%) & 95 ($-65$\%) \\
         & MoS$_2$  & 367 & 385 (+5\%) & 197 ($-46$\%) \\
MACE-H   & Graphene & 266 & 282 (+6\%) & 120 ($-55$\%) \\
         & MoS$_2$  & 379 & 399 (+5\%) & 232 ($-39$\%) \\
\bottomrule
\end{tabular}%
}
\end{table}

\section{Physical constraints of the predicted operators}
\label{app:physical-constraints}

\begin{sloppypar}
The diagnostic suite in \texttt{src/analysis/physical\_constraints.py} tests whether predicted operators satisfy the constraints of electronic-structure theory: positive definiteness of the overlap, occupation bounds and electron count of the density matrix, and consistency between density matrix and Hamiltonian. Table~\ref{tab:physical-constraints} applies it to the validation structures of the three models, after Hermitian symmetrization and restriction to the reference sparsity pattern, and to the DFT reference. Only the silicon model predicts $H$, $S$, and $D$; the SiO$_2$ and ZnCu$_2$Sn(SeS)$_2$ models predict $H$ and are tested together with the reference $S$ and $D$.
\end{sloppypar}

\begin{table}[H]\centering
\caption{Physical-constraint diagnostics on the validation structures, means over $k$-points. $F=\tilde S^{-1/2}H\tilde S^{-1/2}$, $P=\tilde S^{1/2}D\tilde S^{1/2}$, where $\tilde S(\mathbf k)=U\max(\Lambda,10^{-6})U^\dagger$ is the overlap after an eigenvalue floor. Commutation: weight of the off-diagonal part of $P$ in the eigenbasis of $F$, which vanishes if $[F,P]=0$. Fermi--Dirac residual: pooled over the $k$-points of each structure, with $\mu$ fixed by $N_e$ and $T=300$\,K.}
\label{tab:physical-constraints}
\resizebox{\linewidth}{!}{%
\begin{tabular}{llll}
\toprule
Test & Definition & Reference & Prediction \\
\midrule
\multicolumn{4}{l}{\textit{Silicon, predicted $H$, $S$, $D$}}\\
$S$ element error & $\langle|S^{\mathrm{pred}}_{ij}-S^{\mathrm{ref}}_{ij}|\rangle$ & -- & $1.39\times10^{-5}$ \\
$S$ positive definite, as predicted & $\lambda_{\min}(S(\mathbf k))>0$ & 100\% & 17\% \\
$S$ positive definite, after floor & $\lambda_{\min}(\tilde S(\mathbf k))>0$ & 100\% & 100\% \\
Overlap correction & $\|\tilde S-S\|_F/\|S\|_F$ & -- & $4.6\times10^{-5}$ (max $4.0\times10^{-4}$) \\
Condition number & $\kappa(\tilde S)=\lambda_{\max}/\lambda_{\min}$ & $5.4\times10^{3}$ & $5.3\times10^{6}$ \\
Occupation range & $n_i\in\mathrm{spec}(P)$, $0\le n_i\le2$ & $[-0.033,2.30]$ & $[-0.034,2.31]$ \\
Occupations outside $[0,2]$ & $\#\{n_i<0\}$, $\#\{n_i>2\}$ per $k$ & 37.9, 7.0 & 37.6, 7.0 \\
Electron count & $|\mathrm{Tr}(DS)-N_e|$, $N_e=32$ & $1\times10^{-7}$ & $7.4\times10^{-3}$ (max $1.1\times10^{-2}$) \\
Idempotency ($T=0$) & $\|P^2-2P\|_F/2\|P\|_F$ & 0.102 & 0.102 \\
Commutation & $\|P-\mathrm{diag}_F(P)\|_F/\|P\|_F$ & 0.078 & 0.079 \\
Fermi--Dirac residual & $\|P-2f_T(F-\mu)\|_F/\|2f_T(F-\mu)\|_F$ & 0.121 & 0.122 \\
\midrule
\multicolumn{4}{l}{\textit{SiO$_2$, predicted $H$ (10 structures)}}\\
Commutation & as above & 0.063 & 0.127 \\
Fermi--Dirac residual & as above & 0.064 & 0.208 \\
\midrule
\multicolumn{4}{l}{\textit{ZnCu$_2$Sn(SeS)$_2$, predicted $H$ (20 structures)}}\\
Commutation & as above & 0.042 & 0.253 \\
Fermi--Dirac residual & as above & 0.046 & 0.837 \\
\bottomrule
\end{tabular}%
}
\end{table}

The DFT reference itself does not satisfy several of these conditions exactly: the matrices are stored only within a finite cutoff, so $P$ has occupations slightly outside $[0,2]$, and the 300\,K Fermi--Dirac smearing makes exact idempotency inappropriate. The silicon model reproduces these features of its training data to within 2\%. The predicted overlap is positive definite at only 17\% of the $k$-points; an eigenvalue floor at $10^{-6}$ restores positive definiteness with a relative change of $4.6\times10^{-5}$ but leaves the overlap poorly conditioned. This inaccuracy has no practical consequence: $S$ depends only on the atomic positions and the basis set, and the DFT code computes it exactly at negligible cost, without a self-consistent calculation. Predicting $S$ is therefore never necessary, and the spectral results in Section~\ref{sec:results} use the exact overlap. For the Hamiltonian-only models, matrix-element errors of 2--3\,meV are enough to change which states are occupied, most strongly for the small-gap ZnCu$_2$Sn(SeS)$_2$.

\section{Reproducing the demonstration models}
\label{app:reproduction}

\begin{sloppypar}
The training workflows of the three demonstration models are part of the repository in \texttt{scripts/reproduce\_examples}. For each material, a launcher (\texttt{train\_\{zncusnses,sio2,silicon\}.sh}) trains the model from scratch with a frozen JSON recipe that fixes the architecture, the optimizer and learning-rate schedule, the random seeds, and the data split; a README describes their use. The documentation\footnote{\url{https://mandala-org.github.io/Mandala/reproduction/index.html}} explains how to train, continue, and evaluate the models. The datasets and the trained models are available on RODARE (\url{https://doi.org/10.14278/rodare.4816}). Because the workflows are maintained in the repository and run with the current code, they are kept up to date as the software evolves.
\end{sloppypar}

\begin{sloppypar}
The models reported in this work were trained with an earlier version of the FiLM conditioning layer, which modulated features with signed pseudoscalars. These networks are therefore equivariant under rotations but not exactly under inversion. The layer has since been corrected, and new models are E(3)-equivariant. Loading the published models with the option \texttt{film\_semantics: legacy\_2026\_07} reproduces their original predictions; the evaluation script \path{scripts/report/evaluate_validation_error_distribution.py} uses this option and confirms the published validation errors to within $10^{-4}$ relative.
\end{sloppypar}

\end{document}